\documentstyle[aps,prb,epsf,eqsecnum,floats]{revtex}
\begin{document}
\twocolumn[\hsize\textwidth\columnwidth\hsize\csname@twocolumnfalse%
\endcsname

\title{Competing orders and quantum criticality in doped antiferromagnets}
\author{Matthias Vojta, Ying Zhang, and Subir Sachdev}
\address{Department of Physics, Yale University, P.O.
Box 208120, New Haven, CT 06520-8120, USA}
\date{March 10, 2000;
cond-mat/0003163
}

\maketitle

\begin{abstract}
We use a number of large-$N$ limits to explore the competition
between ground states of square lattice doped antiferromagnets
which break electromagnetic U(1), time-reversal, or square lattice
space group symmetries. Among the states we find are $d$-,
$(s^{*}+id)$-, and $(d_{x^2 - y^2} + i d_{xy})$-wave
superconductors, Wigner crystals, Wigner crystals of hole pairs,
orbital antiferromagnets (or staggered-flux states), and states
with spin-Peierls and bond-centered charge stripe order. In the
vicinity of second-order quantum phase transitions between the
states, we go beyond the large-$N$ limit by identifying the
universal quantum field theories for the critical points, and
computing the finite temperature, quantum-critical damping of
fermion spectral functions. We identify candidate critical points
for the recently observed quantum-critical behavior in
photoemission experiments on ${\rm Bi}_2 {\rm Sr}_2 {\rm Ca Cu}_2
{\rm O}_{8+\delta}$ by Valla {\em et al.} (Science {\bf 285}, 2110
(1999)). These involve onset of a charge density wave, or of
broken time-reversal symmetry with $d_{x^2 - y^2} + i d_{xy}$ or
$s^{\ast} + id$ pairing, in a $d$-wave superconductor.
It is not required (although it is allowed) that the stable state in the
doped cuprates to be anything other than
the $d$-wave superconductor---the other states need only be stable
nearby in parameter space.
At finite temperatures, fluctuations associated with these nearby
states lead to the observed fermion damping in the vicinity
of the nodal points in the Brillouin zone. The cases with
broken time-reversal symmetry are appealing because the order
parameter is not required to satisfy any special commensurability
conditions. The observed absence of inelastic damping of
quasiparticles with momenta $(\pi,k)$, $(k, \pi)$ (with $0 \leq k
\leq \pi$) also appears very naturally for the case of fluctuations
to $d_{x^2 - y^2} + i d_{xy}$ order.
\end{abstract}
\pacs{PACS numbers:}

]

\tableofcontents


\section{Introduction}
\label{intro}

A minimal approach to the physics of the cuprate superconductors
is to assume that all relevant ground states can be completely
characterized by the manner in which they break the global symmetries of
the underlying Hamiltonian. Related ideas have been discussed by others in
Refs.~\onlinecite{zaanen2,bondsite,pnas}; for a review by one of us
see Ref.~\onlinecite{sciencereview}, and for early work we
shall extend in this paper see Ref.~\onlinecite{sr}.
The global symmetries are:
\newline
({\em i}) ${\cal S}$ - the electromagnetic U(1) symmetry, which is
broken by the appearance of superconducting order;
\newline
({\em ii}) ${\cal M}$ - the SU(2) spin rotation invariance
symmetry, which is broken by magnetically ordered states like the
N\'{e}el state;
\newline
({\em iii}) ${\cal C}$ - the space group of the square lattice,
which we will consider broken when an observable invariant under
${\cal S}$ and ${\cal M} $, like site or bond-charge density,
is not invariant under space
group transformations; and
\newline
({\em iv}) ${\cal T}$ - time-reversal symmetry.
\newline
Even in this limited approach, a surprisingly rich number of
phases and phase transitions are possible. While at low enough temperatures, every
phase is amenable to a conventional quasiparticle-like
description, anomalous behavior can appear in the vicinity of
second-order quantum critical points between the phases.

This paper will study the competition between phases which break
one or more of ${\cal S}$, ${\cal C}$, and ${\cal T}$
symmetries, and describe the universal theories of
the associated quantum critical points.
We will do this in the context of a number of large-$N$
computations which, by construction, only produce ground states in
which ${\cal M}$ symmetry is preserved. Transitions at which
${\cal M}$ symmetry is broken are quite important for certain
aspects of the physics of the cuprates (as we shall discuss
later in this section), and the absence of
explicit computations for where such transitions may occur is the
primary limitation of our approach.

A brief discussion of mainly the large-$N$ results has appeared in
an earlier paper \cite{stripeprl}.

\begin{figure}[t!]
\epsfxsize=3.5in
\centerline{\epsffile{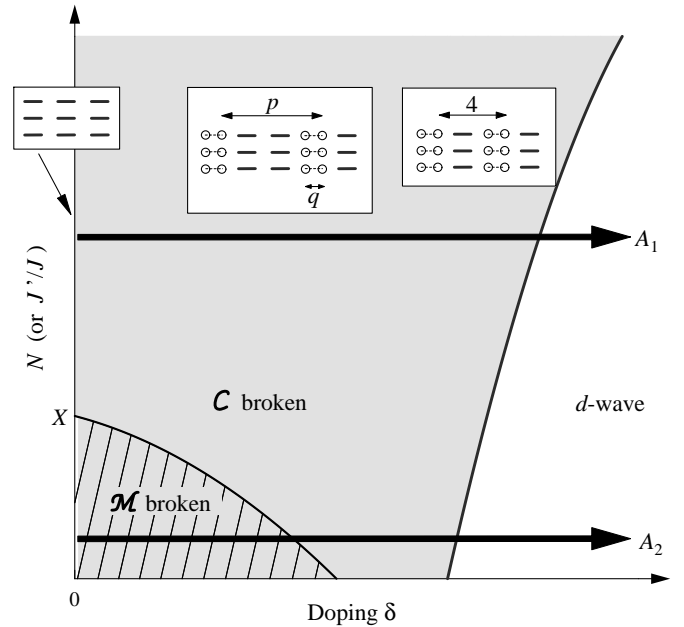}}
\caption{Schematic phase diagram of doped antiferromagnets.
See Section~\protect\ref{intro} for further general discussion and
Section~\protect\ref{sec:spn} for specific quantitative phase
diagrams along the line $A_1$.
The vertical axis represents the size, $N$, of the symmetry group of
spin rotations, ${\cal M}$. Although this parameter is not
experimentally variable, we propose that a similar phase diagram
would be obtained as a function of the ratio of the second ($J'$)
to first ($J$) neighbor exchange interactions.
There is evidence \protect\cite{rs0,gsh,rs1,sr,kotov,sushkov}
that the Peierls order shown above $X$ at
$\delta=0$ is also found in $J^{\prime}-J$ antiferromagnets.
The magnetic ${\cal M}$ symmetry is broken in the
hatched region, while ${\cal C}$ symmetry is broken
in the shaded region; there are numerous additional phase transitions at which
the detailed nature of the ${\cal M}$ or ${\cal C}$ symmetry breaking changes - these
are not shown. For $\delta=0$, ${\cal M}$ symmetry is broken
only below the critical point $X$, while ${\cal C}$ symmetry is broken only
above $X$. Over a significant parameter regime, and for not too
small $\delta$, the ${\cal C}$ symmetry breaking appears in the
stripe patterns shown, with accompanying anisotropic
superconductivity which breaks ${\cal S}$. For some other parameter
regimes (as in Fig~\protect\ref{figmf6}),
the ${\cal C}$ symmetry breaking is realized by
orbital antiferromagnetic (or staggered flux) order: such
${\cal C}$ symmetry breaking cannot survive all the way down to
the point $X$.
The superconductivity is pure $d$-wave only in the large $\delta$ region where
${\cal C}$ and ${\cal M}$ are not broken. The smaller $\delta$
region of the superconductor which preserves ${\cal C}$ and ${\cal M}$
can also exhibit $(d_{x^2-y^2}+ i d_{xy})$- or $(s^{\ast} + id)$-wave superconductivity.
}
\label{fig1}
\end{figure}

A global perspective on our approach is provided by the schematic
phase diagram in Fig.~\ref{fig1} (see also the discussion by
Zaanen\cite{zaanen2}
on related phases).
This sketches the qualitative evolution of the physics as a
function of the hole doping, $\delta$, and $N$, the size of the
symmetry group of spin rotations, ${\cal M}$: we will primarily
consider models in which this symmetry group is generalized from
SU(2) to Sp(2$N$) (specific details in Section~\ref{sec:spn}).

Let us first discuss the physics at $\delta=0$ as a function of
$N$. For small $N$, including the physical case, $N=1$, we know
that the ground state has magnetic N\'{e}el order, and so breaks
${\cal M}$ symmetry. This symmetry is restored by a continuous
quantum phase transition at the point $X$. Above $X$, it is
believed that one enters a paramagnetic phase which generically
has `Peierls' order\cite{rs0,gsh,rs1,sr,kotov,sushkov}:
in such a state all sites are equivalent, but
the charge and energy densities on the bonds\cite{fn1} have the modulation
indicated schematically by the pattern in Fig.~\ref{fig1}. It is
evident that such a state breaks only a ${\cal C}$ symmetry. So
the ${\cal M}$ and ${\cal C}$ symmetries vanish at the
common point $X$, a phenomenon not generically expected in a
Landau theory-like approach suitable for high dimensions, but
possible in the present low-dimensional system with strong quantum
fluctuations \cite{book}. Further above the point $X$, other phases like the
`orbital antiferromagnet' \cite{affmar,schulzoaf,biq,wang,nerses}
(to be described shortly) are also
possible in certain models and we will also discuss these.

The primary purpose of this paper is to study the physics for
$\delta > 0$. Ideally, we should do this along the path $A_2$ in
Fig.~\ref{fig1}, which meets the $\delta=0$ line below $X$.
Instead, we will offer a controlled, quantitative theory
along the path $A_1$ which meets the $\delta=0$ line above $X$.
The implicit assumption underlying such a strategy is that the
quantum critical point $X$ is ``close'' to the regime of
physically relevant parameters. Then we can expect that the
phases with ${\cal M}$ symmetry preserved
along the path $A_1$ are related
to the phases with ${\cal M}$ symmetry preserved
accessed by increasing $\delta$ along the path $A_2$.

This paper describes the intricate interplay between
${\cal C}$, ${\cal S}$, and ${\cal T}$ symmetry breaking
along $A_1$: the body of the paper contains a large number of
phase diagrams as a function of $\delta$ and a dimensionless
measure of the strength of the long-range Coulomb interactions,
and all of these lie along $A_1$. Over a significant regime of
parameters, we find that ${\cal C}$ symmetry is broken at
smaller values of $\delta$; for very small values of $\delta$, the
${\cal C}$-broken phase is an insulating Wigner
crystal-like state, but for larger $\delta$ we obtain a state
with co-existing \cite{zaanen2,kfe} \emph{stripe}
charge order \cite{zaanen,schulz,machida}
and superconductivity, as sketched in Fig.~\ref{fig1}.
Moving to smaller values of $N$ in
Fig.~\ref{fig1}, we expect a transition to a state with
${\cal M}$ symmetry broken which is not contained in our
computations here. The magnetic order appears in a background
of charge stripe order that is present on both sides of the
transition, and the spin polarization is therefore expected to be
incommensurate and \emph{collinear} \cite{bondsite}. It is important to contrast
this magnetic order from the incommensurate ``spiral'' states that
were considered some time ago \cite{shraiman}:
such states have coplanar spins and
do not require co-existing charge order. In contrast, the
incommensurate, collinear spin states must coexist with charge
stripe order \cite{zachar}, and there is evidence that the spin
incommensuration observed experimentally is indeed of this type
\cite{collinear,shirane,younglee}.
Stripe charge order has been discussed first in the context of mean-field
theories for the Hubbard model \cite{zaanen,schulz,machida},
a brief comparison of these results with the present theory
will be given in in Sec.~\ref{secstripes}.

The order parameter for the collinear spin ordering, discussed
above, is an ordinary $O(3)$ vector \cite{zachar,castro,qimp},
identical to that for the transition at the point $X$ at
$\delta=0$. To understand the transition at which ${\cal M}$
symmetry is restored at $\delta > 0$, we need to explore the
possibility of the magnetic order parameter coupling to gapless
fermion excitations. In all our charge stripe states, we find a
strong pairing tendency between the holes, and as a consequence,
the fermion excitation spectrum is either fully gapped, or has
gapless excitations only at special points in the Brillouin zone.
Assuming that momentum conservation prohibits the coupling between
the magnetic order parameter and the gapless fermion excitations
(if present), we arrive at the conclusion that the spin
disordering transition at $\delta > 0 $ is in precisely the same
universality class as that at $X$. Fig.~\ref{fig1} contains a
line, emerging from the point $X$, along which a transition to
${\cal M}$ symmetry restoration takes place. The gist of our
arguments above is that the universality class of the transition
all along this line is likely to be identical to that at $X$. The
implications for such a scenario for experiments
(especially NMR \cite{imai}) has
been reviewed recently in Ref.~\onlinecite{sciencereview}. A
related scenario, and quantitative comparisons with experiments,
has been provided recently by Morr {\em et al.} \cite{morr}.

Returning to the physics along $A_1$, we briefly catalog the
properties of the states found. Further
details appear in Section~\ref{sec:spn}, but the reader is urged
to glance at the phase diagrams in Figs~\ref{figmf1}-\ref{figmf6} at
this stage. It is also worth noting explicitly here that all of
these phase diagrams were obtained in the large $N$ limit, and the
precise numerical values of the parameters at the phase boundaries are not
expected to be accurate for $SU(2)$: nevertheless, the general
topology of the phase diagrams, and the trends in stability
between the various phases, are expected to be realistic.
\newline
({\em i\/}) \underline{\emph{Superconductors with ${\cal C}$ symmetry:}}
These appear for
large $\delta$, and for a large region of parameters
the Cooper pairs are in a
$d$-wave state. We also find a $(s^{*}+id)$-wave state \cite{gabi,sr}
a $(d_{x^2-y^2} + i d_{xy})$-wave state \cite{did1,did2}, both of which
break
${\cal T}$ symmetry, but preserve ${\cal C}$ symmetry;
the latter state
has a non-vanishing spin Hall conductance \cite{senthil}.
\newline
({\em ii\/}) \underline{\emph{Spin-Peierls:}}
An insulating spin-Peierls state at $\delta =0$ was discussed
above: it breaks only ${\cal C}$ symmetry. We also obtained
\cite{sr} for a range of $\delta > 0$ a superconducting
state with precisely the same pattern of ${\cal C}$ symmetry
breaking; naturally, ${\cal S}$ is also broken in such a state.
The signature of both states in neutron scattering would be the
same: all sites are equivalent, but there is a modulation in the
energy and charge densities on the bonds with a period of 2
lattice spacings.
\newline
({\em iii\/}) \underline{\emph{Stripes:}} The striped states are similar to
the superconducting spin-Peierls states above, but all sites are
no longer equivalent. The states have a $p \times 1$ unit cell and
the holes are concentrated on a strip of width $q$; both $p$ and $q$
were always found to be even. The width $q$ regions form strong
one-dimensional superconductors (Luther-Emery liquids), and
coupling between these hole-rich regions leads to anisotropic
superconductivity. The hole pairing also always prefers stripes in a
\emph{bond-centered} configuration \cite{ws,stripeprl}:
the ground state possesses
a reflection symmetry about the centers of certain columns on
bonds, but not about any column of sites. Distinguishing \cite{bondsite}
bond- and site-centering \cite{zaanen,schulz,machida}
is an important issue to be resolved by future
experiments.
\newline
({\em iv\/}) \underline{\emph{Wigner crystal:}} An insulating state for $\delta > 0$
is the familiar Wigner crystal of holes, which breaks only ${\cal C}$
symmetry. This state appears when the strength of the Coulomb
interactions is large enough.
\newline
({\em v\/}) \underline{\emph{Pair crystal:}} An alternative insulating state
can appear at small $\delta>0$ and weaker Coulomb interactions.
The exchange interactions induce pairing of holes, and the
resulting composites then form a Wigner crystal to minimize the
Coulomb repulsion. The underlying square lattice can induce strong
distortions on the usual triangular structure of the Wigner
crystal, so that the state can look like a striped configuration
with an additional longitudinal charge modulation along each hole-rich
stripe.
\newline
({\em vi\/}) \underline{\emph{Orbital antiferromagnet:}} This also known as
the ``staggered flux'' state \cite{affmar,schulzoaf,biq,wang,nerses}.
There are staggered, circulating currents around
each square lattice plaquette, and the state breaks ${\cal T}$
symmetry. The unit cell has two sites and so ${\cal C}$
symmetry is also broken, but a combination of translation by one site
and time-reversal remains unbroken. At half-filling, this state has
gapless fermionic excitations at nodal points along the diagonals
of the Brillouin zone, like the $d$-wave superconductor.
The state with co-existing orbital antiferromagnetism and $d$-wave
superconductivity has broken ${\cal C}$ symmetry (the unit cell
has 2 sites) and is distinct from the $(s^{*}+id)$-wave or
$(d_{x^2-y^2}+i d_{xy})$-wave superconductor.

The competition between ${\cal C}$, ${\cal M}$ and ${\cal S}$
breaking leading to phases with co-existing orders (as
in ({\em ii\/}) and ({\em iii\/}) above) has been discussed by
Zaanen \cite{zaanen2} on phenomenological grounds. However, he
focuses mainly on the bosonic order parameters, while fermionic
excitations will play an important role in our considerations.

Phases related to those in ({\em i\/}), ({\em iii\/}), ({\em
iv\/}) and ({\em v}), and associated phase diagrams, appear
in the work of Kivelson, Fradkin and Emery~\cite{kfe}.
These authors use liquid-crystal like pictures, in which
quasi-independent one-dimensional Luttinger liquids are allowed to
fluctuate transversely, and obtain qualitative phase diagrams.
In contrast, our work is intrinsically
two-dimensional, and the quantum fluctuations are tied more
strongly to the underlying lattice sites; further we shall obtain
quantitative results for phase diagrams, albeit in a large $N$
limit.

It is interesting that none of the states above is an
ordinary Fermi liquid: such a state appears to be always unstable
to some ordering induced by breaking one or more of ${\cal C}$,
${\cal S}$, ${\cal M}$, or ${\cal T}$ symmetries. In this
respect our results are similar to recent results of
Honerkamp {\em et al.} \cite{honer} and Ledermann {\em et al.}
\cite{leder}; however, they find renormalization group instabilities of the
Fermi liquid to states somewhat different from those discussed above.

We conclude our discussion of mean-field theory by noting that a
separate study of ${\cal C}$ symmetry breaking in doped
antiferromagnets has been carried out recently by Stojkovi\'{c}
{\em et al.} \cite{stojk}: they examined the competition between
stripe and Wigner crystal-like phases in a semiclassical theory of
hole dynamics.

\subsection{Quantum phase transitions}
\label{sec:qpt1}

We have already discussed above the nature of the ${\cal M}$
ordering transition in Fig.~\ref{fig1}, and its possible
relationship to NMR experiments \cite{sciencereview,imai}. Here we shall
explore the nature of the quantum phase transitions between the
phases found along $A_1$ (which do not involve order parameters associated with
${\cal M}$ symmetry), and their possible relationship to
quantum critical scaling observed in a recent photoemission
experiment \cite{valla}.

\begin{figure}[t!]
\epsfxsize=3.5in
\centerline{\epsffile{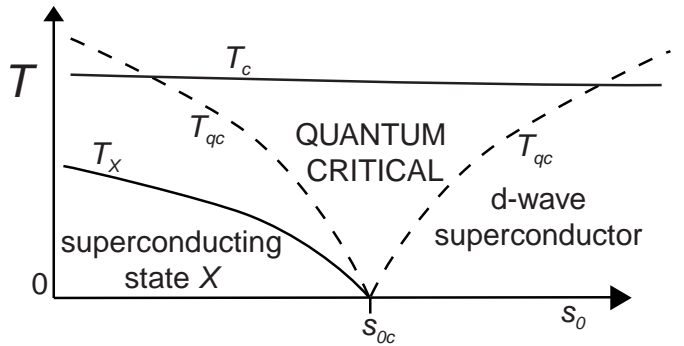}}
\caption{
Finite temperature ($T$) phase diagram in the vicinity of a second
order quantum phase transition from a $d$-wave superconductor as a
function of some parameter in the Hamiltonian,  $s_0$ (which is possibly, but not
necessarily, the hole concentration $\delta$).
Superconductivity is present at temperatures below $T_c$, and the
superfluid density is non-zero on both sides of $s_{0c}$.
The state $X$ is characterized by some other order parameter (in
addition to superconductivity) which vanishes above a temperature
$T_X$. We will consider a number of possibilities for the state
$X$ in this paper, including broken ${\cal T}$ symmetry in
a $(s^{*}+id)$-wave
or a $(d_{x^2-y^2} + i d_{xy})$-wave superconductor, or in an orbital
antiferromagnet, and states with broken ${\cal C}$ symmetry
with charge density wave order. Of particular interest will be
transitions for which the scaling form (\protect\ref{i1}) applies
to the nodal quasiparticles in the quantum critical region $ T >
T_{qc}$. As we discuss in
Section~\protect\ref{sec:qpt1}, this scaling could continue to
apply even above $T_c$ provided the thermal length associated
with proximity to the quantum phase transition at $s=s_{0c}$
remains smaller than the phase coherence length. It is the
proposal of this paper that the high-temperature superconductor
studied in the experiments of Ref~\protect\onlinecite{valla}
is in the vicinity of $s_{0c}$. Indeed, this system could have
$s_0 > s_{0c}$, so that the ultimate ground state is an ordinary
$d$-wave superconductor---the fermion spectrum then exhibits
consequences of fluctuations into state $X$ at $T>T_{qc}$.
}
\label{figx}
\end{figure}

Many of the transitions in our phase diagrams are first order.
These do not have interesting fluctuation spectra, and we will not
consider them further.
We will consider second order transitions at $\delta > 0$, in
which the ground state is superconducting on both sides of the
transitions (see Fig~\ref{figx}). One of these states is a
$d$-wave superconductor, while the other is denoted as
superconducting state $X$ in Fig~\ref{figx}; we will discuss
different candidates for state $X$ below.
Such second order transitions fall into two
classes depending on the behavior of the fermion spectra in the
vicinity of the nodal points of the $d$-wave superconductor. These
points are at \cite{valla} $(\pm K, \pm K)$ with $K=0.391 \pi$ (at optimal doping),
and throughout the remainder of this paper, unless noted otherwise,
we will be implicitly
referring to the fermionic Bogoliubov quasiparticles in the
vicinity of these points. (We will discuss the properties of the
gapped fermionic quasiparticles near $(0,\pi)$, $(\pi, 0)$ at the
end of Section~\ref{intro}.) The two classes are:
\newline
({\bf A}) There is efficient
scattering and damping of the nodal fermionic quasiparticles, and as a
result the fermionic spectral function obeys `naive' quantum
critical scaling (see (\ref{i1}) below),
of the type observed experimentally \cite{valla}.
\newline
({\bf B}) the gapless,
Bogoliubov, fermionic quasiparticles can be neglected in the scaling limit
of the critical theory, and so their damping appears only upon
considering corrections to scaling, and vanishes with a
super-linear powers of temperature ($T$) as $T \rightarrow 0$.

The simplest of the transitions in class A are those that involve
time-reversal symmetry breaking in the $d$-wave superconductor:
transitions from a $d$-wave superconductor to a state $X$ which is
either a $(s^{*}+id)$-wave \cite{gabi,sr}
or a $(d_{x^2-y^2} + i d_{xy})$-wave \cite{did1,did2} superconductor.
It is important to note that both these transitions occur for only
at a {\em finite} attractive coupling in the $s^{\ast}$ or
$d_{xy}$ pairing channels. This is to be contrasted to pairing
instabilities of a Fermi liquid, which would occur at
infinitesimal attraction in either channel. However, when the
parent state is a $d_{x^2-y^2}$-wave superconductor, the vanishing
density of states at the Fermi level removes the usual BCS log
divergence in the Cooper pair propagator, and a finite attraction is
required for further pairing in the $s^{\ast}$ or
$d_{xy}$ channels. This finite coupling instability is directly
responsible for a non-trivial quantum critical point, with strong
thermal and quantum fluctuations leading to class A behavior,
whose effects we shall describe
and exploit in this paper.
Much of the more recent discussion (see Ref.~\onlinecite{clap}
and references therein)
 of $(d_{x^2-y^2} + i d_{xy})$-wave
superconductivity has focused on the case where this order is
induced by external perturbations like an applied magnetic field,
and is motivated partly by the experiments of
Ref.~\onlinecite{didexp}. In contrast, in our paper, we are
interested in the spontaneous appearance of such order, and this
is necessarily associated with a sharp transition and ${\cal T}$ symmetry breaking.
There is a divergent susceptibility associated with this symmetry
breaking, and this will lead to a low energy amplitude fluctuation
mode distinct from that discussed in Ref.~\onlinecite{clap}.

Another transition involving breaking of time-reversal symmetry is
that between a $d$-wave superconductor and a state $X$
in which $d$-wave superconductivity and orbital antiferromagnetism
coexist. Unlike the above, this transition will be shown to be in class B.

A slightly more complicated transition in class A is one in which
$X$ involves the
onset of ${\cal C}$ symmetry breaking in a $d$-wave superconductor: such a
transition is the boundary of the ${\cal C}$-broken region in
Fig.~\ref{fig1}. However, a special condition is required for such a
transition to be in class A: the charge-ordering wavevector should
precisely equal the wavevector between two nodal points in the
$d$-wave superconductor; otherwise the transition is in class B.
The theory for such a transition is closely related to models for
the onset of antiferromagnetism in a $d$-wave superconductor
considered recently by Balents {\em et al.} \cite{balents}.
We note that others have also discussed quantum phase transitions
involving stripe or charge density wave
order in the cuprate superconductors in recent years\cite{zaanen2,castellani};
however, in contrast to us,
these works have either ignored interplay with the dynamic
properties of the fermions\cite{zaanen2}, or focused on transitions
in a Fermi liquid ground state\cite{castellani}, not a
$d$-wave superconductor.

The hallmark of the class A transition is that in its $T>0$
``quantum-critical'' region \cite{book} (see Fig~\ref{figx}),
the fermion Green's function near one of the nodal points
of the $d$-wave superconductor obeys
\begin{equation}
G_f ( k , \omega) = \frac{{\cal A}_f}{T^{(1-\eta_f)/z}} \Phi_f \left(
\frac{\hbar \omega}{k_B T}, \frac{v k}{(k_B T)^{1/z}} \right);
\label{i1}
\end{equation}
here $z$ is the dynamic critical exponent ($z=1$ for the specific models solved
in this paper), $k$ measures the distance from one of the nodal points of
the $d$-wave superconductor, $\omega$ is a measuring
frequency, $v$ is a velocity (for $z=1$), ${\cal A}$
is an overall amplitude, $\eta_f$ is a universal anomalous dimension,
and $\Phi_f$ is a universal scaling function of its two arguments.
We emphasize that unless the system happens to be precisely at the
quantum-critical point, which is generically not expected to be
the case, the scaling form (\ref{i1}) will eventually fail as $T
\rightarrow 0$. Indeed, if define a lower crossover
temperature $T_{qc}$ so that (\ref{i1}) holds for $T>T_{qc}$,
then $T_{qc} \sim |s_0 - s_{0c}|^{z \nu}$ where $s_0$ is some
coupling constant in the Hamiltonian, the $T=0$ quantum critical
point is at $s_0= s_{0c}$, and $\nu$ is the usual correlation length critical
exponent. For $T< T_{qc}$, normal Bogoliubov
quasiparticle behavior emerges, and this is indeed observed
\cite{valla,shen,campu,fed,mesot,mesot2} to be the case experimentally
at very low $T$. We emphasize further that it is
not even necessary that the point $s_0 = s_{0c}$ be in an
experimentally accessible parameter regime. So if we are
considering the $d$-wave to $(d_{x^2-y^2} + i d_{xy})$-wave transition (for
definiteness), then it is not required (although it is permissible)
that the true ground state
of a cuprate compound be a $(d_{x^2-y^2} + i d_{xy})$-wave superconductor
over some doping regime. It is only necessary that a $(d_{x^2-y^2} + i d_{xy})$-wave
superconductor be close enough to the physical regime, so
that dynamic fluctuations to $(d_{x^2-y^2} + i d_{xy})$-wave order
are apparent in the quantum-critical regime. Postulating the
existence of a $(d_{x^2-y^2} + i d_{xy})$-wave ground state somewhere in
parameter space is then a powerful theoretical tool for obtaining
a controlled description of this intermediate temperature regime.
When this paper was almost complete,
new experimental evidence for broken ${\cal T}$
symmetry near defects was reported \cite{kirtley};
these results support the hypothesis that the
bulk energy of a superconductor with broken ${\cal T}$ symmetry
is not very much higher than that of a $d$-wave superconductor \cite{sigrist},
and that a quantum phase transition between these states may indeed be
near the experimentally accessible parameter space.

One of the purposes of this paper is to develop a method for
computing the universal function $\Phi_f$ in (\ref{i1})
for the various transitions in
class A noted above. We will find that two different methods are
necessary, depending on the frequency/wavevector regime being
accessed. For $v k \gg k_B T$ or $\hbar \omega \gg k_B T$, a
straightforward resummation of a renormalized perturbative
expansion suffices. However, for $v k < k_B T$ and $\hbar \omega < k_B
T$, an entirely different approach has to be developed. Now
there is strong damping induced by scattering between thermally
induced excitations, and we compute it in a self-consistent theory
of excitations scattering via a renormalized, temperature
dependent $T$-matrix.

It is important to keep in mind that the class A transitions being
considered here have long-range superconducting order on both
sides of the quantum critical point. The order parameter
associated with the transition involves either ${\cal C}$ or ${\cal T}$
symmetry breaking, and has no direct relationship to ${\cal S}$
symmetry. The main role of the superconducting order is to define
the bare spectrum of the fermion excitations which then interact
with the critical order parameter fluctuations. In the experiments
\cite{valla}, scaling related to (\ref{i1}) is also observed above
the superconducting transition temperature, $T_c$ (which is quite distinct from
$T_{qc}$ and could be either above or below it); indeed there is no
signature of $T_c$ in the photoemission spectrum (while below $T_{qc}$
there is a crossover to conventional quasiparticle behavior). Our
quantum-critical theory entirely neglects the fluctuations of the
superconducting order itself, but this does not limit its
applicability to below $T_c$; rather, we only need to impose the more
limited constraint that the phase coherence length of the
superconducting order parameter is larger than the inelastic
scattering length of the class A transition with ${\cal C}$ or ${\cal T}$
symmetry breaking. This constraint is automatically satisfied
below $T_c$, and can easily be satisfied over a wide range of
temperatures above $T_c$. Indeed, the latter length, by
(\ref{i1}), decreases as $\sim 1/T^{1/z}$, and so the
constraint
becomes less stringent as $T$ increases.

Finally, we discuss the important issue of the gapped
quasiparticles well away
from the nodal points of the $d$-wave superconductor.
The fermionic quasiparticles with momenta $(\pi,k)$, $(k, \pi)$
(with $0 \leq k \leq \pi$)
have a non-zero excitation energy which
is a minimum near \cite{mesot,mesot2} $k \approx 0.18 \pi$ at optimal doping.
Experiments \cite{shen,campu,fed} clearly
indicate that these quasiparticles are
sharply defined at all temperatures below $T_c$, and do not show
any sign of ``quantum-critical'' damping (we thank M.~Norman and M.~Randeria for
emphasizing this to us). So the order parameter fluctuations discussed above,
which
are responsible for the quantum-critical damping near the nodal
points $(\pm K,\pm K)$, clearly cannot couple efficiently to
the quasiparticles on the lines between $(\pi,\pi)$ and
$(\pi,0)$, $(0,\pi)$.
For the case of a transition from $d$-wave to $(d_{x^2-y^2} + i
d_{xy})$-wave superconductivity (discussed in Section~\ref{sec:dsid})
this is, in fact, very naturally
the case: the $d_{xy}$ order parameter $\sim \sin k_x \sin k_y$
vanishes when either $k_x=\pi$ or $k_y=\pi$.
For the transition from $d$-wave to $(s^{*}+id)$-wave (also
discussed in Section~\ref{sec:dsid}),
the $s^{\ast}$ order parameter $\sim (\cos k_x + \cos k_y)$,
and this vanishes at the points $(\pi,0)$, $(0, \pi)$;
however there will be some
residual coupling as one moves away from these points to
$(\pi,k)$, $(k,\pi)$ with $k \approx 0.18 \pi$.
Finally, for the transitions in class A
involving the onset of ${\cal C}$ symmetry breaking (to be
discussed in Section~\ref{sec:cd}), momentum conservation
makes the coupling of the order parameter to fermions
near $(\pi,0)$, $(0, \pi)$ very ineffective: the order parameter
scatters the fermions to a region of the Brillouin zone where the
quasiparticles have an even higher energy.

The outline of the remainder of this paper is as follows. In
Section~\ref{sec:spn} we present the results of the large-$N$
study along $A_1$. The universal theories of the second-order
quantum phase transitions appear in Section~\ref{sec:qpt}.
A summary of our results and a discussion of experimental issues
is in Section~\ref{sec:conc}. A calculation of the fermion
damping in a naive renormalized perturbation theory, and its
failure in the low frequency regime $\hbar \omega < k_B T$ is
discussed in Appendix~\ref{app:damp}.
Readers not interested in specific details of our results
can glance at Figs~\ref{figmf1}-\ref{figmf6} and move ahead to
Section~\ref{sec:conc}.


\section{S\lowercase{p}(2$N$) \lowercase{$t$}-$J$ model in the large-$N$ limit}
\label{sec:spn}

For a microscopic investigation of the ground states of doped antiferromagnets
we start from the usual $t-J$ model ${\cal H}_{tJ}$ on the sites $i$ of a square lattice,
which is complemented by a Coulomb ${\cal H}_V$ interaction between
the electrons, ${\cal H} = {\cal H}_{tJ} + {\cal H}_{V}$,
\begin{eqnarray}
{\cal H}_{tJ} = \sum_{i > j}
&& \left[ -t_{ij} {c}_{i \sigma}^{\dagger}
{c}_{j\sigma} + {\rm H.c.}  + J_{ij} \left( {\bf S}_i \cdot {\bf S}_j -
\frac{{n}_i {n}_j}{4} \right) \right] \,,\nonumber \\
{\cal H}_{V} = \sum_{i > j} && V_{ij} {n}_i {n}_j .
\label{e1}
\end{eqnarray}
The electron operators $c^\dagger$ exclude double occupancies
which expresses the (infinite) electronic on-site repulsion.
${n}_i = {c}_{i\sigma}^{\dagger} {c}_{i\sigma}$ is the
charge density at site $i$.
We will primarily concerned with the case where the fermion
hopping, $t_{ij}$, and exchange, $J_{ij}$, act only when $i,j$
are nearest neighbors, in which case $t_{ij} = t$ and $J_{ij} =
J$; however, we will occasionally refer to cases with second
neighbor hopping ($t'$) or exchange ($J'$).
For the off-site Coulomb repulsion $V_{ij}$ we assume
weak or no screening since the zero-temperature ground states
will be either insulating or superconducting; in both
cases the density of states at the Fermi level vanishes.
Therefore we will use a $1/R$ decay of the interaction,
$V_{ij} = V / |{\bf R}_i-{\bf R}_j|$,
and the strength of the repulsion is parameterized by $V$.
The $V_{ij}$ are included to counter-act the
phase separation tendency of the $t-J$
model~\cite{ekl,sr,sk1,sk2,hm}, and play a key role in our analysis.

We shall be interested in describing the ground state of ${\cal H}$
as a function of its couplings and the average doping
concentration $\delta$.
We generalize the spin symmetry \cite{rs1,sr,sstri} from SU(2) to Sp(2$N$) and examine
the limit of large degeneracy $N$.
In the large-$N$ approach the ground state of the system can be found
in a saddle-point approximation (which becomes exact for $N=\infty$).
Large-$N$ expansions have been applied to a large number of
antiferromagnetic spin systems as well as to models with doping.
The motivation for using the symplectic Sp(2$N$) generalization instead
of the more common SU($N$) variant is that it does not rely on a
two-sublattice structure of the underlying antiferromagnet and is
therefore more appropriate for systems where frustration may play a
role \cite{rs1}.
Furthermore, the Sp(2$N$) approach includes naturally pairing
of spins which leads to superconducting ground states when the system
is doped.
(Note that both Sp(2$N$) and SU(2$N$) reduce to the
usual SU(2) symmetry group for $N=1$.)
Apart from studies of $t$-$J$ models like (\ref{e1}), Hubbard
models with Sp(2$N$) symmetry have also been
studied \cite{ziqiang,foerster}.

The behavior of the system in the large-$N$ limit depends on the representation
of Sp($2N$) used for the spin operators.
In the context of SU($N$) approaches to the 2D Heisenberg model
especially totally symmetric (bosonic) and totally antisymmetric (fermionic)
representations have been applied, see e.g. Refs. \onlinecite{affmar,biq,rsnp,grilli}.
For the present problem of the $t-J$ model it turns out that a simple
large-$N$ limit (leading to a saddle point in the free-energy functional)
exists only for a fermionic representation of the spin degrees of
freedom.
As discussed in the introduction,
in this case the ground state of the undoped (Heisenberg) model does not
break spin rotation symmetry $\cal M$.
Instead, translation and rotation symmetries $\cal C$ are broken
and the state has been shown to be
a paramagnetic spin-Peierls state which can
also be considered as a {\em bond-centered} charge density wave.
Recent work~\cite{kotov} has shown strong evidence for this order in the
{\em frustrated} SU(2) quantum antiferromagnet on a square lattice
(with $J' > 0$).

Let us now describe the details of the large-$N$ approach.
In the following we consider spins transforming under the antisymmetric product of
$m$ fundamentals of Sp(2$N$), the large-$N$ limit is taken with $m/N$ constant.
The spins are represented by fermions $f^\alpha$, $\alpha=1 \ldots 2N$, which
transform under the fundamental of Sp(2$N$).
The holes are described by spinless bosons $b$,
$c_i^{\alpha} = f_i^{\alpha} b_i^\dagger$.
The local constraint of the $t-J$ model acquires the form
\begin{equation}
f_{i\alpha}^\dagger f_i^{\alpha} + b_i^\dagger b_i = m
\,.
\end{equation}
Here we will only discuss states being half-filled at zero doping, $m=N$.
The average hole concentration $\delta$ is determined by
\begin{equation}
\frac{1}{N_s} \sum_i f_{i\alpha}^\dagger f_i^{\alpha} = N(1-\delta)
\end{equation}
where $N_s$ denotes the (infinite) number of lattice sites.
Within the Sp(2$N$) generalization of the system,
the spin operators ${\bf S}_i$ become fermion bilinears times the traceless
generators of Sp(2$N$),
the Hamiltonian (\ref{e1}) takes the form
\begin{eqnarray}
{\cal H}_{t} = \sum_{i > j} &&
\left[ -\,\frac{t_{ij}}{N} \, b_i f_{i\alpha}^\dagger f_j^{\alpha} b_j^\dagger + {\rm H.c.}
  \right ] \nonumber \\
+ \sum_i && \lambda_i \left(f_{i\alpha}^\dagger f_i^{\alpha} + b_i^\dagger b_i - N \right)
+ \mu \sum_i \left( b_i^\dagger b_i - N\delta \right)
  \,, \nonumber \\
{\cal H}_{J} =  \sum_{i > j} &&
 \left [ - \frac{J_{ij}}{2N}
\left({\cal J}^{\alpha\beta} f_{i\alpha}^{\dagger} f_{j \beta}^{\dagger} \right)
\left({\cal J}_{\gamma\delta} f_{j}^{\delta} f_{i}^{\gamma} \right)
  \right] \nonumber \\
{\cal H}_{V} = \sum_{i > j} &&
\frac{V_{ij}}{N} b_i^\dagger b_i b_j^\dagger b_j
\label{e2}
\end{eqnarray}
where we have split ${\cal H}_{tJ} = {\cal H}_{t} + {\cal H}_{J}$ for
convenience.
The Lagrange multipliers $\lambda_i$ enforce the local occupation constraint, and
$\mu$ fixes the average hole density.
${\cal J}^{\alpha\beta}$ denotes the antisymmetric Sp(2$N$) tensor:
\begin{equation}
{\cal J}^{\alpha\beta} = {\cal J}_{\alpha\beta} =
\left(
\begin{array}{cccccc}
   &1&  & &  & \\
-1&  &  &  &  & \\
   &  &  & 1&  &  \\
 & &-1& & & \\
 & &  & &\ddots& \\
 & &  & & &\ddots\\
\end{array}
\right)
\end{equation}
We remind the reader that $\cal H$ given in eq. (\ref{e2}) reduces
to eq. (\ref{e1}) for the ``physical'' case of $N=1$.

In the limit $N=\infty$ at zero temperature the bosons $b_i$ condense,
$\langle b_i \rangle = \sqrt{N} b_i$, so $b_i^2$ is the hole density
and $N(1-b_i^2) = \langle {n}_i \rangle$ the charge density at site $i$.
The exclusion of double occupancies in the $N=1$ case is now represented
by the on-site constraint $\langle {n}_i\rangle \leq N$ since
$b_i^2 >0$.
The long-range nature of Coulomb interaction requires the introduction
of a background charge of magnitude $\delta$ on each lattice site
for the total system to be charge-neutral.
The interaction is decoupled by the introduction of the link fields $Q_{ij}$
with the saddle-point values
\begin{equation}
N Q_{ij} =
\langle {\cal J}^{\alpha\beta} f_{i\alpha}^{\dagger}
          f_{j \beta}^{\dagger} \rangle
=
\frac{1}{b_i b_j}
\langle {\cal J}^{\alpha\beta} c_{i\alpha}^{\dagger}
          c_{j \beta}^{\dagger} \rangle
\,.
\label{intdecoup}
\end{equation}
The $Q_{ij}$ represent the complex bond pairing amplitudes.
We note that $Q_{ij} = Q_{ji}$, and the phases of the
$Q_{ij}$ are only fixed up to a global gauge
transformation, $f_i \to f_i e^{i \Theta}$, which leads
to $Q_{ij} \to Q_{ij} e^{-2 i \Theta}$.
However, the plaquette operator $\Pi_Q$, which can be defined as
$\Pi_Q = Q_{12} Q_{23}^{*} Q_{34} Q_{41}^{*}$
for the sites 1...4 at the corners of a unit square,
is a gauge-invariant object.

At the saddle point the Hamiltonian takes the form
\begin{eqnarray}
{\cal H}_{t} = \sum_{i > j} &&
\left[ -\, t_{ij} \, b_i b_j f_{i\alpha}^\dagger f_j^{\alpha}  + {\rm H.c.}
  \right ]  \nonumber \\
+ \sum_i && \lambda_i \left(f_{i\alpha}^\dagger f_i^{\alpha} + N b_i^2 - N \right)
+ \mu N \sum_i \left( b_i^2 - \delta \right)
  \,, \nonumber \\
{\cal H}_{J} = \sum_{i > j} &&
 \left[ -\frac{J_{ij}}{2} \left({\cal J}^{\alpha\beta} f_{i\alpha}^{\dagger} f_{j \beta}^{\dagger}
       Q_{ij}^{*} + {\rm H.c.} - N |Q_{ij}|^2 \right)
  \right] \nonumber \\
{\cal H}_{V} = \sum_{i > j} &&
N \,V_{ij} (b_i^2-\delta) (b_j^2-\delta)
\,.
\label{e3}
\end{eqnarray}
which is bilinear in the fermions and can be solved by
a Bogoliubov transformation.
The saddle point solution is found by minimizing the
total free energy with respect to $b_i$ and $Q_{ij}$
at fixed average fermion occupation and hole
density.
The saddle-point equations have been solved numerically with
unit cell sizes up to 32 sites.
The Coulomb repulsion term in the large-$N$ limit is purely ``classical'', i.e.,
it does not involve the fermions.
Using the lattice Fourier transforms $V_{\bf k}$ and $n_{\bf k}$ of interaction
and charge distribution respectively,
the Coulomb contribution to the energy can be re-written as
\begin{eqnarray}
{\cal H}_{V} = N_s N \sum_{\bf k} V_{\bf k} n_{\bf k} n_{-\bf k}
\label{ev}
\end{eqnarray}
The interaction $V_{\bf k} = V \sum_{{\bf R}\neq 0} e^{{\rm i}{\bf k R}}/|{\bf R}|$ behaves
as $\sim k^{-1}$ for small $k=|{\bf k}|$, but becomes negative around the center of the Brillouin
zone.
(Note that there is no on-site contribution from $V$, $V_{ii}=0$, since the occupation
constraint is already taken care of by the local chemical potential $\lambda_i$.)


The numerical determination of the minimum energy configuration in the
large-$N$ limit consists of two nested loops:
(i) Starting from an initial guess for $Q_{ij}$, $b_i$, $\lambda_i$
the fermionic Hamiltonian is diagonalized using a discrete momentum
grid ($32^2$ is sufficient for unit cells up to 8 sites).
The expectation values $\langle f_{i\alpha}^{\dagger} f_{j \beta}^{\dagger} \rangle$
provide new values for the $Q_{ij}$, new $b_i$ are obtained from
$b_i^2= 1 - \langle f_{i\alpha}^\dagger f_i^{\alpha} \rangle / N$,
and the average chemical is adjusted (by a simple bisection step) to match
the doping level. This is repeated until convergence is reached.
(ii) The optimum charge distribution within the unit cell of $N_c$ sites is found
by minimizing the total energy [obtained in loop (i)] w.r.t. the
differences in the chemical potentials $\tilde\lambda_i = \lambda_i-\lambda_1$.
This is a minimum search in a $(N_c-1)$-dimensional space and can be performed
by standard methods.
To account for the possible existence of more than one saddle point
the initial link field values $Q_{ij}$ are chosen randomly, and several sets
of initial conditions are used to identify the saddle point corresponding
to the {\em global} minimum energy.


\subsection{Ground states at $N=\infty$}

The results of our large-$N$ calculation can be summarized as
follows:
First, at $\delta =0$ along $A_1$ we find the fully dimerized,
insulating
spin-Peierls (or $2 \times 1$ bond charge density wave)
solution~\cite{rsnp}
in which $|Q_{ij}|$ is non-zero only on the bonds shown in
Fig.~\ref{fig1}.

At non-zero doping $\delta$ the ``bare'' large-$N$ $t-J$ model shows phase
separation for a large range of parameters $t/J$,
see Ref. \onlinecite{sr}.
This tendency to phase separation (into a hole-rich region and a
fully dimerized half-filled region) at $V=0$ is an important
ingredient of our study.
With the inclusion of ${\cal H}_V$ the phase separation becomes
``frustrated'', as emphasized in Refs.~\onlinecite{ekl,sk1,sk2},
and the competition of the energy scales $t$, $J$, and $V$ leads
to various kinds of charge ordering phenomena.
Different ground state phases may be realized depending on the
strength on the Coulomb interaction.
The details of these solutions will be described in the
next subsections.

\begin{figure}
\epsfxsize=3.2in
\centerline{\epsffile{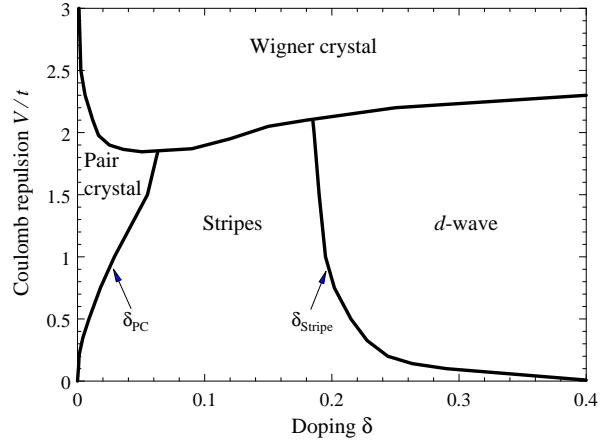}}
\caption{
Ground state phase diagram of $\cal H$ at
$N=\infty$, $t/J=1.25$.
Except for the $d$-wave superconductor all states
have $\cal C$ broken.
The ``stripe'' states have coexisting charge density wave order
and superconductivity; the crystal states are
insulating.
The thick lines denote phase transitions being first-order in the
large-$N$ limit.
Within the charge-ordered phases there are numerous additional transitions at which
the detailed nature of the ${\cal C}$ symmetry breaking changes - these
are not shown.
The left and right boundaries of the stripe phase define
$\delta_{\rm PC}(V)$ and $\delta_{\rm Stripe}(V)$, see text.
}
\label{figmf1}
\end{figure}

Fig.~\ref{figmf1} shows a representative phase diagram containing a
cut along the $V$--$\delta$ plane of the parameter space for fixed $t/J=1.25$.
Phase diagrams for other parameter values are shown in
Figs.~\ref{figmf2} and \ref{figmf3}.
The inclusion of additional model parameters as a biquadratic exchange
interaction can lead to further ground state phases, these
will be described in Sec.~\ref{secaddpara}.

Let us start the discussion of Fig.~\ref{figmf1} with large doping. Here
no phase separation tendency is present, the ground
state charge distribution is homogeneous. The magnetic interaction
together with the (infinite) on-site repulsion leads to pairing
in the $d_{x^2-y^2}$ channel.
Moving to smaller doping, the system encounters a phase separation
instability at $V=0$, i.e., the holes
are expelled because the magnetic interaction favors a fully dimerized
half-filled configuration.
The Coulomb repulsion counteracts (frustrates) this phase
separation tendency \cite{ekl} leading to
microscopic charge ordering below a certain hole concentration,
$\delta < \delta_{\rm Stripe}$.
The important point now is that the kinetic energy disfavors crystal-like
states where the holes are essentially localized.
Instead, the holes form ``stripes'' where hopping in one dimension is
still possible.
Furthermore, the tendencies to fermion pairing on one hand
and to dimerization on the other hand are still present.
Consistent with this picture, our numerical search always yielded lowest energy
states with ${\cal C}$ and $\cal S$ broken, in a
region $\delta_{\rm PC} < \delta < \delta_{\rm Stripe}$ and
for small values of the Coulomb repulsion $V$.
These states consist of {\em bond-centered charge density waves}~\cite{ws,stripeprl}
(stripes) which co-exist with superconductivity.
The effect of the kinetic energy becomes less pronounced when the doping is
further decreased.
Eventually, for extremely small doping, $\delta < \delta_{\rm PC}(t,J,V)$, and
non-zero repulsion $V$ the lowest energy state is an insulating crystal of Cooper
pairs which breaks $\cal C$ symmetry.
Such a state arises from the combination of charge ordering and pairing tendencies.
Depending upon the parameter values $t$, $t'$, $J$ and $V$ also
a doped spin-Peierls state may be realized at small and/or intermediate
doping.
For strong Coulomb repulsion between the charges (large $V$ region in Fig.~\ref{figmf1})
one expects that hopping as well as pairing (mediated by the magnetic interaction)
will become unimportant.
Then the situation resembles a low-density electron gas (where the
potential energy dominates over the kinetic energy), and
the ground state becomes a Wigner crystal built of single charges.

We continue with a more detailed description of the characteristics of
the mean-field ground states.

\subsubsection{Homogeneous superconductor}

At doping levels smaller than a critical $\delta_{\rm PS}(t,J)$
the ``bare'' $t-J$ model in the large-$N$ limit shows phase separation \cite{sr}
as explained above. [Note $\delta_{\rm PS}(t,J) = \delta_{\rm Stripe}(t,J,V=0)$.]
In contrast, for $\delta > \delta_{\rm PS}$ and small $V$ the ground states have
uniform site charge distributions.
These states include a doped spin-Peierls state ($2\times 1$ unit cell, see below)
at relatively small $\delta$, and states with a single site per unit cell
which are homogeneous superconductors.
They can be characterized by two link field values $Q_x$ and $Q_y$.
In particular, one finds a $d$-wave superconductor with $Q_x = -Q_y$ at
intermediate $\delta$ ($\sim$ 20-50\%),
and an extended $s^*$-wave superconductor with $Q_x=Q_y$ for larger $\delta$.
The excitation spectrum of the $d$-wave superconductor has four nodes
whereas the $s^*$-wave phase is fully gapped.

Note that for $t/J < 0.3$ the critical doping level
$\delta_{\rm PS}$ is in fact zero, so ${\cal H}_{tJ}$ does not show phase
separation (and no stripes for $V>0$) at very small $t/J$.
Also, in the regime of small $t/J$ the $d$-wave superconductor is replaced
by a state with $s^* + i d_{x^2-y^2}$ symmetry, i.e., the link fields obey
$Q_y = Q_x e^{i \theta}$ with a continuously varying phase $\theta$.
Here the quasiparticle spectrum is again fully gapped, furthermore
time-reversal symmetry $\cal T$ is broken.

The described results for the homogeneous states are identical to the ones
obtained in Ref.~\onlinecite{sr}.
Similar states are also found for large values of $|t'/t|$
where the tendency to phase separation is suppressed,
see Sec.~\ref{secaddpara}.

\subsubsection{Spin-Peierls state}
\label{secsp}

The spin-Peierls state which breaks $\cal C$ is found as ground state of
the undoped system in the large-$N$ limit.
For $\delta=0$ its energy per site is given by $E_{\rm SP} / (N N_s) = -J/4$.
Only one of the four link fields of the $2\times 1$ unit cell is non-zero, i.e.,
the square lattice is completely covered by dimers.

The corresponding doped state is the lowest-energy state with a {\it homogeneous}
site-charge distribution in the small doping regime\cite{sr}.
The link fields in $x$ direction $Q_{1x}, Q_{2x}$ have different magnitudes reflecting
the dimerization, whereas $Q_{1y} = Q_{2y}$.
The fields cannot be made real simultaneously in any gauge, so the doped state
breaks time-reversal invariance $\cal T$ and is fully gapped.
However, if we restrict our attention to states
which preserve $\cal T$, then there exists a region of small dimerization,
$|Q_{1x} - Q_{2x}| \ll |Q_{1x}|$,
near the transition to a $d$-wave state where the spectrum has 4 gapless
points.
Note that this state which has coexisting superconducting and
spin-Peierls order can be considered as a $2 \times 1$ bond
charge density wave with the hole densities being equal on
all sites. Related states have been studied in other approaches
\cite{grilli,sushkov}, but they find spin-Peierls order in a Fermi liquid, not
a superconductor.

\subsubsection{States with stripe charge order}
\label{secstripes}

The saddle-point solutions at small doping,
$\delta_{\rm PC} < \delta < \delta_{\rm Stripe}$, and not too large $V$,
break lattice translation symmetry ${\cal C}$
and can be described as bond-centered charge density waves.
These states have a $p \times 1$ unit cell,
as shown in Fig.~\ref{fig1}.
We always found $p$ to be an even integer, reflecting the
dimerization tendency of the $\delta=0$ solution.

\begin{figure}[t!]
\epsfxsize=3.2in
\centerline{\epsffile{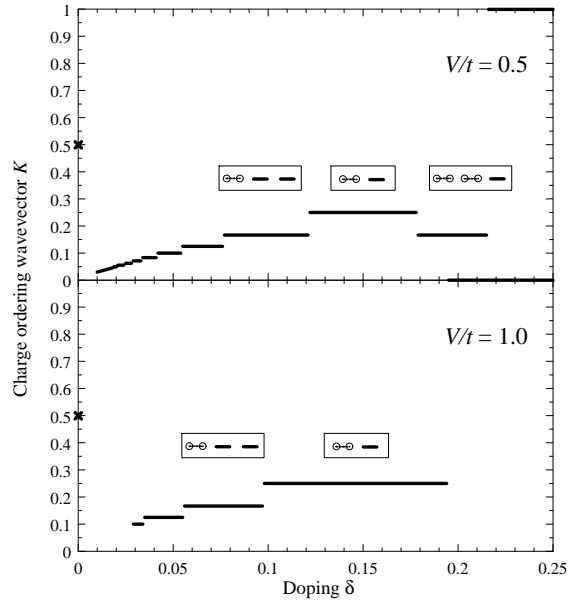}}
\caption{
The charge-ordering wavevector, $K$, (in reciprocal lattice
units) as a function of $\delta$ at $N=\infty$ for
$t/J = 1.25$ (Fig.~\protect\ref{figmf1}),
and two different strengths of the Coulomb
repulsion $V/t=0.5$ and 1.0.
For the stripe states as illustrated in Fig.~\protect\ref{fig1}
we have $K=1/p$,
the spin-Peierls state has $K=1/2$ at $\delta =0$.
The $K=1$ value at large $\delta$ has ${\cal C}$
symmetry restored, and is a pure $d$-wave superconductor.
For very small $\delta$, the ground state is a Wigner crystal of
Cooper pairs with incommensurate charge order.
The unit cells for the largest plateaus are also shown.
}
\label{figmf1b}
\end{figure}

The holes prefer to segregate in one-dimensional striped
structures, i.e., within each $p \times 1$
unit cell, the holes are concentrated on a $q \times 1$ region.
The link fields $Q$ in the hole-rich region can be made real and
have different signs in $x$ and $y$ direction reminiscent of
$d$-wave pairing correlation in the stripes.
Most of the stripe phases are ``fully'' formed stripes,
i.e., the regions between the stripes with a width of $p-q$ have a hole
density of zero and are insulating and fully dimerized.
In this case the link fields between the stripe sites and the
boundary sites of the insulating region vanish.
However, for a parameter region close to the
transition to a doped spin-Peierls state (see Fig.~\ref{figmf2})
there occur ``partially'' formed stripes as ground states.
These states have the same kind of charge density modulation
as described above, but the hole density in the hole-poor region is
not zero, so the system is an anisotropic 2D superconductor
even at $N=\infty$.
In general,
the hole density $\rho_{\ell}$ per unit length of each stripe is not of
order unity as found in earlier theories \cite{} but
significantly smaller.
The values of $q$ and $\rho_{\ell}$ are determined
primarily by $t$, $J$, $V$;
they depend only weakly on $\delta$.
For intermediate values of $V$ where $q=2$ we found values
of $\rho_{\ell} \sim 1/2$, here the stripe can be viewed as a ladder
with roughly 1/4 hole per site.
In general, smaller values of $V$ yield larger values of $q$;
the limit $V \rightarrow 0$ leads to $q\rightarrow \infty$ which
reflects the tendency to phase separation in the ``bare'' $t-J$
model.

The main effect of varying the total hole density is a change of $p$ and
therefore of the stripe distance.
We note that the stripes do not survive in the limit $\delta\to 0$ \cite{coulnote}:
for $\delta < \delta_{\rm PC}$ the ground state changes to an insulating Wigner
crystal of Cooper pairs.
However, for small $V$, $\delta_{\rm PC}$ is very small,
$\delta_{\rm PC} \sim {\rm exp}(-1/V)$.
For $\delta_{\rm PC} \ll \delta \ll \delta_{\rm Stripe}$, we find
an approximate proportionality $p \sim 1/\delta$, which also
implies that $\rho_{\ell}$ is nearly independent of
$\delta$.
The evolution of the ordering wavevector $K=1/p$ with
$\delta$ is shown in Figs.~\ref{figmf1b},
\ref{figmf2}, \ref{figmf3},
there are plateaus for each even integer number $p$.
Our large-$N$ theory only found ``stripe'' states in which $K$
was quantized at the rational plateaus in Fig.~\ref{figmf1b}.
The reason for $p, q$ being even and for the plateaus is easily identified
as the strong dimerization tendency of the system.
The columnar arrangement of the spin-Peierls singlets immediately
leads to the ``staircase''-like curve shown in Figs.~\ref{figmf1b},
\ref{figmf2}, \ref{figmf3}.
However, for
smaller $N$ we expect that irrational, incommensurate, values of $K$ will
appear, and interpolate smoothly between the plateau regions.

For most values of $t$, $J$, and $V$ there is a large plateau at $p=4$
around doping $\delta = 1/8$, and, for some
parameter regimes, this is the last state before ${\cal C}$ is
restored at large $\delta$; indeed $p=4$ is the smallest value of $p$
for which our mean-field theory has solutions with $b_i$ not
spatially uniform. Experimentally~\cite{jtran,birg}, a pinning of the charge
order at a wavevector $K=1/4$ is observed, and we
consider it significant that this value emerges naturally from
our theory.

It is worth pointing out that the hole density $\rho_{\ell}$  per unit length
of stripe is not exactly pinned at one value
which means that the stripes are {\it not} incompressible.
The hole density varies continuously within each plateau
but jumps discontinuously as the transition is made from one
plateau to the next.

There are strong pairing correlations between the holes in each
$q$-width region.
Strictly speaking, for $N=\infty$ each $q$-width stripe above is a
one-dimensional superconductor, while the intervening
$(q-p)$-width regions are insulating. However, fluctuation
corrections will couple with superconducting regions, yielding
an effective theory discussed in Section VII of
Ref.~\onlinecite{mpaf} with their dimensionless parameter $K \sim N$.
This implies that Josephson pair tunneling between the one-dimensional
superconductors is a relevant perturbation at sufficiently large $N$, leading
to a two-dimensional anisotropic superconducting
ground state, and should allow good metallic conduction above the
superconducting transition temperature. These characteristics are
consistent with observations\cite{uchida} on ${\rm La}_{2-x-y}{\rm Nd}_y {\rm
Sr}_x {\rm Cu O}_4$.

\begin{figure}[t!]
\epsfxsize=3.5in
\centerline{\epsffile{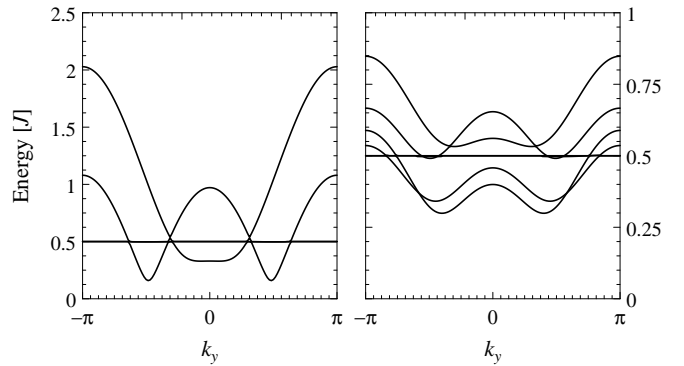}}
\caption{
Fermionic mean-field excitation spectrum in fully striped phases,
$t/J=1.25$.
The spectrum is independent of $k_x$, the momentum perpendicular
to the stripes.
Left: $q=2$ phase with $\rho_{\ell}=0.8$.
Right: $q=4$ phase with $\rho_{\ell}=0.6$.
In the second case the bandwidth is much smaller due to the
smaller doping level in the stripes ($\rho_{\ell}/q$).
The flat bands at energy $J/2$ correspond to excitations
of localized dimers in the undoped regions.
}
\label{figmfspec}
\end{figure}

The quasiparticle spectrum in the striped phases is always fully
gapped because of the presence of superconductivity.
Except for the partially striped phases where the fermion spectrum shows
a weak dependence on the momentum $k_x$ perpendicular to the
stripes \cite{stripeprl}, the spectrum of the stripe phases is $k_x$-independent.
The fermion spectrum in the partially striped phases were
displayed in Ref.~\onlinecite{stripeprl}, and in
Fig.~\ref{figmfspec} we show the quasiparticle bands for the fully striped
cases.
The spectrum consists of two contributions:
The excitations of the fully dimerized undoped regions correspond to
removing one fermion from a localized dimer, the energy cost is $J/2$ and
of course momentum-independent.
The stripes of width $q$ give rise to $q$ one-dimensional dispersing
bands where the gap is determined by the pairing amplitude of order $J$, and
the total bandwidth is given by $4 t \rho_{\ell}/q$ where $\rho_{\ell}$ is
the linear hole density per unit stripe as above.
The stripe excitations therefore lie in the gap of the spin-Peierls state;
their properties are entirely determined by $\rho_{\ell}$ and $q$ (and the model
parameters $t$ and $J$).
Upon increasing the doping level $\delta$ within one of the plateaus
in Fig.~\ref{figmf1b} the superconducting gap decreases slightly since
the link fields $Q$ decrease, and the position of the dispersion minimum
changes corresponding to the Fermi momentum.
The transitions between the plateaus arise from level crossings, i.e.,
there is no critical dynamics associated with them.
At the transition from one plateau to the next, $\rho_{\ell}$ decreases
discontinuously, leading to an increase in the gap and so on.
If the stripe width $q$ changes at a plateau transition then the number
of dispersing bands will also change, see Fig.~\ref{figmfspec}.

\begin{figure}[t!]
\epsfxsize=3.2in
\centerline{\epsffile{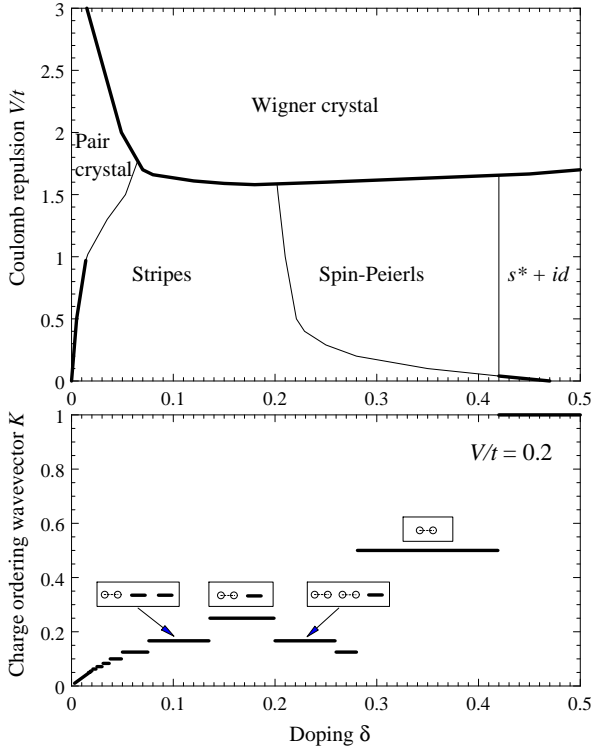}}
\caption{
Same as in Figs.~\protect\ref{figmf1} and \protect\ref{figmf1b},
but for $t/J = 0.5$.
Upper panel: Ground state phases of $\cal H$ as function of doping and
Coulomb repulsion.
Thick and thin lines denote first-order and second-order transitions,
respectively.
Lower panel: Charge-ordering wavevector, $K$, (in reciprocal lattice units)
for $V/t = 0.2$.
For $28\% < \delta < 42\%$ the ground state is the doped spin-Peierls state
breaking $\cal C$ and $\cal S$;
it has a uniform site-charge distribution and a $2\times 1$ unit cell.
}
\label{figmf2}
\end{figure}

At this point we briefly discuss earlier mean-field calculations
\cite{zaanen,schulz,machida} which predicted an inhomgeneous charge
distribution in the ground state of the Hubbard and related
models.
These computations where based on the observation that large-$S$
mean-field theories of doped antiferromagnets
($S$ denotes the size of the spin) show charge density wave
instabilities -- the same applies to the
large-$N$ theory discussed here.
An important difference, however, is the character of the stripes.
The early mean-field calculations for the Hubbard model
\cite{zaanen,schulz,machida} predict insulating, site-centered stripes with a
hole density of unity within the hole-rich regions;
the experimentally found stripe states in the cuprates are,
however, either metallic or superconducting.
In contrast to these early mean-field results, the present large-$N$
computations yield superconducting, bond-centered stripes,
and it would be useful for future experiments to detect
this distinction between bond- and site-centering.
Let us discuss this difference more precisely for the case $p=4$.
For the site-centered stripe, the hole density per
unit length in each column of sites takes values $\rho_1$,
$\rho_2$, $\rho_3$, $\rho_2$ before repeating periodically (where
$\rho_{1-3}$ are some three distinct densities); this is the configuration
usually assumed in most experimental papers.
In contrast, for the
bond-centered state, these densities take the values
$\rho_1$, $\rho_1$, $\rho_2$, $\rho_2$, and this also appears to
be compatible with existing observations.
The bond-centering is important in the present computation
for enhancing pairing correlations, which are responsible for
superconducting/metallic transport in the direction parallel to the
stripes.

Possible differences in the spin dynamics between bond- and site-centered
striped phases have been discussed by Tworzydlo {\em et al.} \cite{bondsite}.
They argued that the magnetic domains between the stripes can be described
by spin ladders with either an even or odd number of legs; these two cases
lead to very different spin fluctuation spectra.

We note that a very recent NQR experiment\cite{julich} indicates
a charge distribution which has some features
consistent with the bond-centered state: they find only two
inequivalent sites associated with the stripes, and a density in
the hole-rich region which is considerably smaller than 0.5.

\begin{figure}[t!]
\epsfxsize=3.2in
\centerline{\epsffile{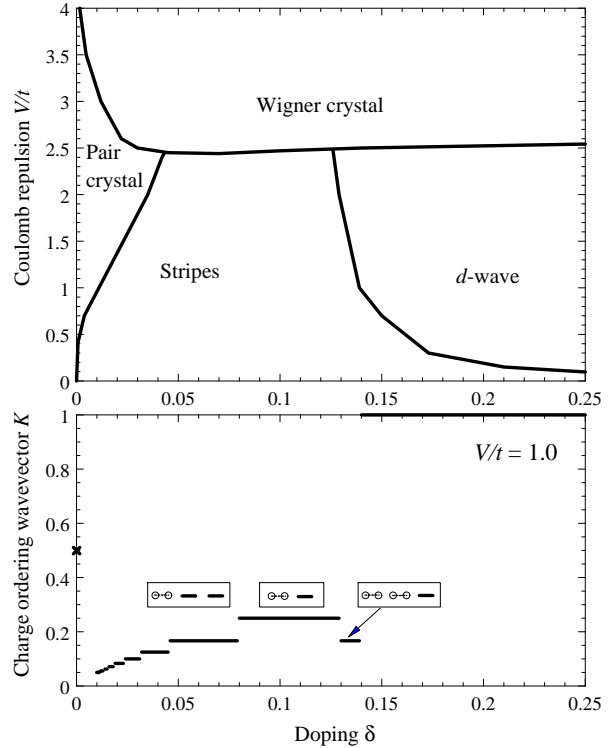}}
\caption{
Same as in Figs.~\protect\ref{figmf1} and \protect\ref{figmf1b},
but for $t/J = 2.5$.
Upper panel: Ground state phases of $\cal H$ as function of doping and
Coulomb repulsion.
Lower panel: Charge-ordering wavevector, $K$, (in reciprocal lattice units)
for $V/t = 1.0$.
}
\label{figmf3}
\end{figure}

\subsubsection{Wigner crystals of Cooper pairs}

For very small doping the ``stripe'' states become unstable
with respect to further ${\cal C}$ breaking.
This leads to two-dimensional insulating states with ${\cal S}$
restored, which can be characterized as Wigner crystals of Cooper
pairs \cite{ekl}.
%
%
In the large-$N$ limit, the Cooper pair crystals consist of pairs of adjacent
sites with non-zero hole density and a non-zero bond amplitude $Q$ between them;
the remaining sites of the square lattice form singlet bonds as in the
fully dimerized, undoped spin-Peierls state.
The occurrence of such pair crystal states is intimately related to the
dimerization tendency of the system:
In a pair crystal the repulsion energy is much smaller than a stripe state,
and the hole-rich sites still form dimers leading to a non-zero contribution
from ${\cal H}_{tJ}$.
(At $N=\infty$ the hole density on the sites of each pair is not exactly
unity, $b_i^2<1$, this is an artifact of the large-$N$ limit.
For physical values of $N$ the physical system will certainly have two
electrons per pair.)

To minimize the repulsion energy, the hole pairs want to form a triangular
Wigner lattice (with a lattice constant $\sim \sqrt{\delta}$).
However, in the present problem the hole positions have
to be on the underlying square lattice. This leads
to incommensurate structures
where the pairs form an approximate triangular lattice.
The energy per site in this state at small $\delta$
is given by $E_{\rm PC}/(N N_s) = -J/4 - \alpha\,\delta$ where
$\alpha$ contains contributions from $t$, $J$, and $V$.

We note here that there are also crystal-like insulating saddle-point
solutions with clusters of sites with non-zero hole density.
These ``charge islands'' can be considered as an integer number
of Cooper pairs bound together -- the described pair crystals
are the smallest possible islands.
Larger islands occur at small doping, they replace the pair crystal phases
shown in the phase diagrams at small hopping $t$.
We have not systematically examined all possible ``island'' phases
because large islands require large unit cells in the numerical
calculation.
However, by estimating the ground state energy of the ``island'' phases
we have checked that stripe phases are always favored in the experimentally
interesting doping regime, $8\% < \delta < 20 \%$.

Summarizing these findings,
the frustration of complete phase separation by $V$ leads to two possible
scenarios of charge clusters with ``smaller'' size:
(i) stripes which are 1d objects, i.e., the charges are confined in
one direction (note that finite $T$ may melt the stripe order
leaving behind finite length parts of stripes; strong impurity influence
may also break up stripes into segments), and
(ii) islands which can be considered as bound states of Cooper pairs --
here the charges are confined in both directions.
The calculations show that for physical values of $t/J$ and not too small
doping $\delta$ the stripe scenario (i) is favored due
to the larger kinetic energy of the holes in the stripes.

\subsubsection{Wigner crystals of single charges}

For large $V$ the Coulomb repulsion dominates over $t$ and $J$
which leads to the appearance of Wigner crystal
states -- here the crystal consists of {\em single charges}.
The holes minimize the Coulomb energy by forming a crystal-like
structure -- this is similar to the charge ordering in a low-density
electron gas.
The lowest Coulomb energy is again reached for the case where the charges form a
triangular lattice with a Wigner lattice constant $\sim \sqrt{\delta}$, however,
the hole positions have to coincide with the sites of
the underlying square lattice.
Therefore, the states at large Coulomb repulsion $V$ will have structures
incommensurate with the square lattice, i.e., a (infinitely) large unit cell
with holes sitting on a fraction $1/\delta$ of the sites ($b_i^2=1$).
These sites form an approximate triangular Wigner lattice, all other sites
are half-filled ($b_i^2=0$) and form dimerized bonds.
(The calculation shows that at finite values of $V/J$ the competition of
exchange and Coulomb energies at $N=\infty$ leads to hole densities on the
sites of the Wigner lattice being smaller than unity, $b_i^2<1$, and
to a distortion of the ideal dimer pattern in the link fields $Q$ near
each hole site.
This again is an artifact of the condensation of the slave bosons in the
large-$N$ limit.)

The kinetic energy in the Wigner crystal state vanishes.
The exchange energy per site
is given by $-J/4 \times (1-\delta)$ at low $\delta$ since $(1-\delta)$ is
the number of sites remaining for singlet formation.
It is important to note that the Coulomb energy is negative, i.e.,
$\langle {\cal H}_V \rangle \sim - \gamma\,\delta^{1/2} < 0$ with
$\gamma \sim V$.
So we can estimate the energy per site for small $\delta$ to be
$E_{\rm WC}/(N N_s) = -J/4 (1-\delta) - \gamma\,\delta^{3/2}$.

\subsection{Influence of additional model parameters}
\label{secaddpara}

The calculations presented so far show that the tendency to
stripe formation at low doping is a very robust feature of
the $t-J-V$ model at $N=\infty$.
In the following we discuss the influence of the inclusion of further
processes into the Hamiltonian.

\subsubsection{Longer-range hopping}
\label{tpr}

We start with an additional next-nearest neighbor hopping $t'$ which is
supposed to be important in certain cuprate superconductors.
The large-$N$ calculations show that the results are nearly independent of $t'$ as
long as it is small, $|t'/t| < 0.3$.
Larger values of $t'$ of either sign suppress the tendency to phase separation in
the ``bare'' model (without ${\cal H}_V$).
This in turn means that the stripe instability is weakened;
the stripe region in the phase diagram shrinks, it is shifted to lower
doping and lower values of $V$.
This is consistent with e.g. the findings of White and Scalapino \cite{ws}.

\subsubsection{Biquadratic exchange}

We have also considered the addition of a biquadratic spin exchange
\begin{equation}
H_4 = \frac{\kappa J}{4 N^3} \sum_{<i,j>} ({\bf S}_i \cdot {\bf S}_j)^2
\end{equation}
to the Hamiltonian (\ref{e2}) for general $N$.
The strength of the biquadratic interaction is given by $\kappa J$.
For the physical case $N=1$ the biquadratic exchange term simply reduces to
${\bf S}_i \cdot {\bf S}_j + {\rm const}$, but for larger $N$ it
contains exchange processes involving eight fermions \cite{biq}.
For the total coupling of two (isolated) spins to be antiferromagnetic
there is an upper limit for $\kappa$ which is given by ${\kappa} < 2$
for any $N$ \cite{biq}.
In the large-$N$ limit the decoupling can be done in two steps,
it turns out that no new link fields are necessary. At the saddle
point the additional term takes the form:
\begin{equation}
H_4 = \frac{\kappa J}{8} \sum_{<i,j>} |Q_{ij}|^2
       \left(2 {\cal J}^{\alpha\beta} f_{i\alpha}^{\dagger} f_{j \beta}^{\dagger}
       Q_{ij}^{*} + {\rm H.c.} - 3 N |Q_{ij}|^2 \right)
\label{e4}
\end{equation}
As discussed in Ref. \onlinecite{biq} such a biquadratic exchange
term weakens the tendency to dimerization.
In fact, in the SU($N$) saddle-point approach it can be used to
stabilize a staggered flux phase;
it is expected to find a similar effect on the hole-poor regions
of the stripe states \cite{bradlec}.

In the undoped system, we find that the $N=\infty$ ground state changes
from spin-Peierls (at ${\kappa}=0$) to a box phase for
$0 < \kappa < 1.13$, and then to a ``flux''-like phase for
$\kappa > 1.45$.
(Some intermediate phase occurs for $1.13 < \kappa < 1.45$.)
All phases at ${J}>0$ can be described within a $2\times 2$ unit
cell.
The box phase has non-zero link fields $Q$ around isolated plaquettes of four
sites each, all other link fields vanish \cite{biq}.
In the ``flux''-like phase all $Q$ are equal in magnitude, their phases can be made
real by a gauge transformation, and the expectation value of the plaquette
operator $\Pi_Q$ is real and negative.
However, unlike in the SU($N$) case there is no direction associated with the
$Q$ link fields.
A direct interpretation in terms of a flux is therefore not possible.
The described mean-field solutions are identical with the results of Marston and
Affleck for the SU($N$) Heisenberg model \cite{biq}.
In fact, by applying a generalized particle-hole transformation it can be
shown that the mean-field equations for the undoped system
are equivalent for both SU(2$N$) and Sp(2$N$) cases.

Turning to finite doping, we find that all homogeneous phases are
unstable w.r.t. phase separation at $V=0$ and low doping.
The inclusion of $V$ leads to stripe states similar to the ones described
in Sec.~\ref{secstripes} for all values of $\kappa$.
The hole concentration $\rho_{\ell}$ in the stripes is again determined
by the microscopic parameters, and is found to be between 0.2 and 0.5 for
reasonable values of $t$, $J$, and $V$.
The stripes are one-dimensional superconductors,
for most of the parameter values the bond amplitudes $Q$ within the stripes
can be made real and reflect $d$-wave pairing correlations
[at larger $V$ and intermediate $\delta$ there is a small region with
$(s^{*}+id)$-like pairing in the stripes].
The $Q$ fields in the insulating regions between the stripes reflect the structure
of the undoped ground state: At small non-zero $\kappa$ one finds a box-like
structure whereas $\kappa > 1.45$ gives rise to a more homogeneous distribution of
the $Q$ which can be understood as flux-like arrangement with distortions at the
boundaries (stripes).
Interestingly, in the latter case there can occur non-zero link fields $Q$ between the
stripes and the undoped region (leading to longer-ranged spin correlations), but
these disappear for larger hole concentration in the stripes.

\begin{figure}[t!]
\epsfxsize=3.2in
\centerline{\epsffile{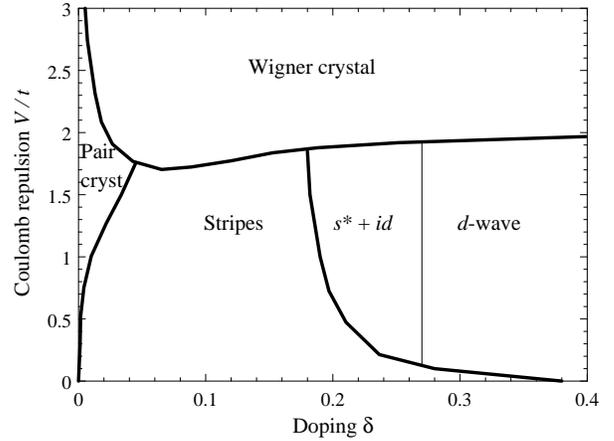}}
\caption{
Ground state phase diagram for the Sp($2N$) $t-J$ model with biquadratic
exchange term as function of doping and Coulomb repulsion,
$t/J = 0.75$, $\kappa=1.5$.
}
\label{figmf4}
\end{figure}

At larger doping the stripes are disfavored, the ground state
becomes a homogeneous superconductor with either $d$-wave or
$s^{*}+id$ symmetry.
Compared to the case $\kappa = 0$, the tendency to phase separation is
slightly weakened by the introduction of the biquadratic exchange,
the region of stripe ground states shrinks.
A sample phase diagram is shown in Fig.~\ref{figmf4}.
Note that the transition from $d$-wave to $s^* + id$ symmetry with decreasing doping is
accompanied with the opening of a gap in the fermionic spectrum (and with
spontaneous $\cal T$ breaking).

\subsubsection{Next-nearest neighbor exchange}
\label{jpr}

To explore further possible symmetries of the superconducting order
parameter (which is represented by the link fields $Q$ in the
mean-field theory) we consider an additional next-nearest neighbor
exchange $J' > 0$.
Note that the inclusion of such a frustrating next-nearest neighbor
exchange between pairs of sites on the same sublattice
is straightforward within the present Sp(2$N$) mean-field theory
since the same representation of Sp(2$N$) is employed for spins
on both sublattices [in contrast to the bosonic SU(N) approach].
The new interaction is decoupled similar to the nearest-neighbor
interaction as described at the beginning of Sec.~\ref{sec:spn};
this leads to diagonal link fields $Q'_{ij}$.
The model with hopping terms $t$ and $t'$, exchange processes $J$ and
$J'$ as well as biquadratic exchange strengths $\kappa J$ and
$\kappa J'$ shows a number of new ground state phases at the
mean-field level, among them superconductors with mixed order
parameters, spin-Peierls and flux-like phases as well as
charge-ordered states.

\begin{figure}[t!]
\epsfxsize=3.2in
\centerline{\epsffile{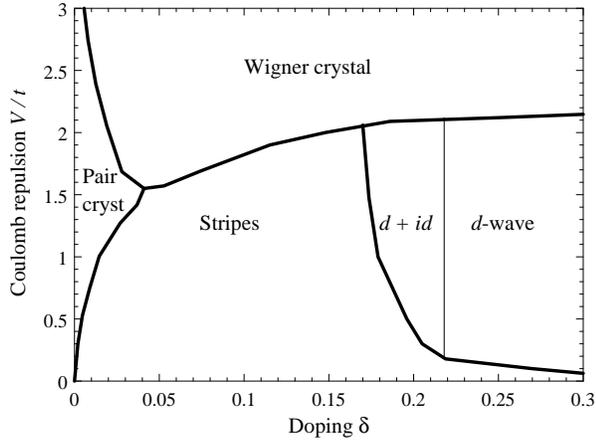}}
\caption{
Ground state phase diagram for the Sp($2N$) $t-J-J'-V$ model
with biquadratic exchange term as function of doping and Coulomb
repulsion; $t/J = 1.0$, $J'/J = 0.4$, $t'=0$, $\kappa=1.5$.
$d+id$ denotes the $\cal T$-breaking superconductor with
$(d_{x^2 - y^2} + i d_{xy})$-wave symmetry.
}
\label{figmf4b}
\end{figure}

As an example, we show the phase diagram for $t/J=1.0$, $J'/J = 0.4$,
and $\kappa = 1.5$ in Fig.~\ref{figmf4b}.
The undoped system has a large ground state degeneracy,
small doping produces stripe states due to the phase separation
instability in the absence of Couplomb repulsion $V$.
The stripes disappear at larger doping, leading to a superconductor
with a homogeneous charge distribution.
First we find a $(d_{x^2 - y^2} + i d_{xy})$-wave
superconductor; further increasing $\delta$ leads
to a pure $d_{x^2 - y^2}$ state.
The $(d_{x^2 - y^2} + i d_{xy})$-wave state has a fully gapped spectrum
and breaks time reversal symmetry.
It is characterized by four $Q$ link fields which obey
$Q_x=-Q_y$, $Q'_{11}=-Q'_{1,-1}$ and $Q'_{11} = i \epsilon Q_x$.
Here, $\epsilon$ is a real value measuring the mixing of
both $d$-wave order parameters.
At the transition line between the $(d_{x^2 - y^2} + i d_{xy})$ state
and the pure $d_{x^2 - y^2}$-wave state
the $\epsilon$ vanishes continuously.

\subsubsection{Alternative interaction decoupling}
\label{mixint}

There has been recent interest \cite{ivanov,nayak} in possible ground states which
spontaneously break time-reversal symmetry $\cal T$ and show
circulating currents \cite{affmar,schulzoaf,biq,wang,nerses} (doped ``flux'' phases).
Such states can be characterized as orbital magnets; if the current
direction around each unit cell alternates in sign between adjacent cells
one obtains an orbital antiferromagnet.
To explore possible realizations of such states within our saddle-point
approach we extend the Hamiltonian by an interaction term which provides decoupling
in the particle-hole channel in the large-$N$ limit.
It has the form $(-{\bar J}_{ij})/(2N)\,(f_{i\alpha}^{\dagger} f_{j\alpha})(f_{j\beta}^{\dagger} f_{i\beta})$
which also reduces to ${\bf S}_i \cdot {\bf S}_j + {\rm const}$ for
$N=1$.
The interaction is decoupled by link fields $\chi$ which take the values
\begin{equation}
N \chi_{ij} = N \chi_{ji}^{*} =
\langle f_{i\alpha}^{\dagger}  f_{j \alpha} \rangle
\label{intdecoup2}
\end{equation}
at the saddle point.
In contrast to the $Q$ fields these new fields $\chi_{ij}$ specify a
direction for the link $(ij)$.
For the undoped case the phases of the $\chi$ are gauge-dependent, but
the plaquette operator $\Pi_{\chi} = \chi_{12} \chi_{23} \chi_{34} \chi_{41}$
is again a gauge-invariant object.
In the doped system, the fields $\chi$ are directly connected with the current
transported through one link, $j_{ij} = 2 N \,{\rm Im} \chi_{ij}$, so their
phases are meaningful.

At the saddle point the Hamiltonian containing the interaction $\bar J$ takes the form
\begin{equation}
\bar{\cal H}_{J} = \sum_{i > j}
 \left[ -\frac{{\bar J}_{ij}}{2} \left( f_{i\alpha}^{\dagger} f_{j \alpha} \chi_{ij}^{*} + {\rm H.c.}
   - N |\chi_{ij}|^2 \right)
  \right]
\label{e3b}
\end{equation}
which is formally a contribution to the fermion hopping.
We also include a corresponding biquadratic interaction which can be decoupled
in the particle-hole channel leading to
\begin{equation}
\bar{H_4} = \frac{\kappa {\bar J}}{8} \sum_{<i,j>} |\chi_{ij}|^2
       \left(2 f_{i\alpha}^{\dagger} f_{j \alpha}
       \chi_{ij}^{*} + {\rm H.c.} - 3 N |\chi_{ij}|^2 \right)
\label{e4b}
\end{equation}
The mean-field equations show that a magnetic interaction of the type (\ref{e3b})
is equivalent to
the decoupling done within a SU(2$N$) large-$N$ approach to the antiferromagnet.
The sum of both terms, ${\cal H}_J + \bar{\cal H}_J$, contains both
possible factorizations (particle-particle and particle-hole) of the
four-fermion interaction term.
It is therefore equivalent to an unrestricted Hartree-Fock treatment
of the original Heisenberg Hamiltonian for $N=1$ written in terms
of auxiliary fermions, provided that $J_{ij} = {\bar J}_{ij}$.
In the following, we will discuss the large-$N$ limit and treat $J$ and $\bar J$
as independent parameters, concentrating on the region where $J \sim \bar J$.
Varying the ratio $x \equiv {\bar J} / ({\bar J} + J)$ can be viewed as
continuous interpolation between the Sp($2N$) ($x=0$) and SU($2N$) ($x=1$) approaches to the
antiferromagnet.

\begin{figure}[t!]
\epsfxsize=3.2in
\centerline{\epsffile{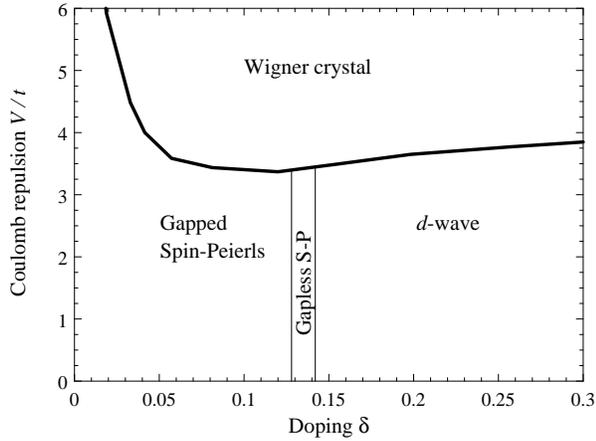}}
\caption{
Ground state phase diagram for the $t-J$ model with interactions
${\cal H}_J$ and $\bar{\cal H}_J$, $t/J=1$, $x=1/2$ (${\bar J}=J$),
$\kappa=0$ (no biquadratic exchange).
There are {\em two} spin-Peierls phases shown:
At larger doping the dimerization is small leading to gapless spectrum with
four nodes.
Upon decreasing $\delta$ the dimerization increases, and when it exceeds some
critical value, the spectrum becomes gapped: the gap opens when
the nodal points collide in pairs at points on the lines $k_y = \pm \pi/2$
in the extended Brillouin zone.
}
\label{figmf5}
\end{figure}

For $x=1/2$ ($J=\bar J$) and doping $\delta=0$, we find a large ground state degeneracy
which has also been reported earlier in unrestricted Hartree-Fock treatment of the
Heisenberg antiferromagnet\cite{KotLiu}.
The degeneracy can be lifted by the introduction of a biquadratic exchange $\kappa>0$
which leads to box-like phases having a $2\times 2$ unit cell.
These states have non-zero values of both the $Q$ and $\chi$ link fields.

Turning to finite doping, we find no phase separation instability at $x=1/2$ independent
of $\kappa$ ---
at zero or small $V$ states with a homogeneous site charge distribution always have
lower energy than inhomogeneous states; there are no stripe states
at $x=1/2$.
The ground state at low doping is either a spin-Peierls state (at $\kappa=0$) or
box-like ($\kappa>0$), for larger doping it becomes a pure $d$-wave superconductor.
The phase diagram for $\kappa=0$ is shown in Fig.~\ref{figmf5}.
The spin-Peierls state shows dimerization in both the $Q$ and the $\chi$
fields with a pattern as described in Sec.~\ref{secsp}.
It is interesting to note that the doped spin-Peierls state does not break time reversal
symmetry $\cal T$, so the $Q$ fields can be made real by a gauge transformation
(in contrast to the spin-Peierls state occurring in the ``pure''
Sp(2$N$) case which is always $\cal T$ breaking).
Similarly, the doped box state for $x=1/2$ also preserves $\cal T$.

\begin{figure}[t!]
\epsfxsize=3.2in
\centerline{\epsffile{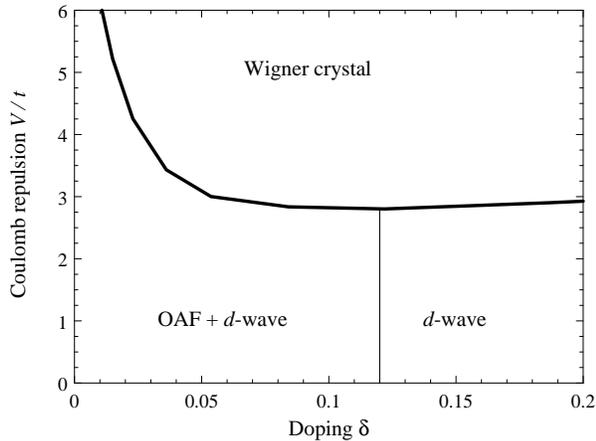}}
\caption{
Ground state phase diagram for the $t-J$ model with interactions
${\cal H}_J$ and $\bar{\cal H}_J$ which are both supplemented
by a biquadratic exchange, $\kappa=1.5$.
The remaining parameters are $t/J=1.25$, $x=0.6$ (${\bar J}= 1.5 J$).
OAF denotes the $\cal T$-breaking orbital antiferromagnet with
staggered circulating currents around each plaquette.
At $\delta=0$ we have the staggered flux state.
}
\label{figmf6}
\end{figure}

As one further example, we describe the ground states at $x=0.6$.
The undoped system without biquadratic exchange ($\kappa=0$) again shows
a large degeneracy.
Values of $0<\kappa<1.4$ lead to a box-like phase.
For $\kappa>1.4$ we find the staggered flux phase as ground state,
here all $\chi$ are equal in magnitude and the plaquette operators
$\Pi_{\chi}$ have the negative real expectation values.
The particle-particle link fields $Q$ vanish for any $\kappa$.

At finite doping, there is again no phase separation tendency at $x=0.6$.
For $\kappa=0$ the ground state phases are spin-Peierls at small $\delta$
and $d$-wave at larger $\delta$, whereas small non-zero $\kappa$
produces box-like ground states.
Interesting phases occur for $\kappa>1.4$ as shown in the phase diagram in
Fig.~\ref{figmf6}:
The staggered flux phase
evolves to an orbital antiferromagnet with a two-site unit cell and
circulating currents (given by ${\rm Im}\,\chi$) in each plaquette with
alternating direction between adjacent plaquettes.
This state is unstable to pairing \cite{wang}:
the $Q$ fields acquire non-zero, real values leading to a $\cal T$-breaking
state with coexisting orbital antiferromagnetism and $d$-wave
superconductivity \cite{nayak}.
We note that this coexistence occurs here at the mean-field level,
this can be contrasted to other mean-field theories \cite{zeyer} where
the flux and $d$-wave instabilities occur in different orders in a $1/N$
expansion.

For larger doping, the currents vanish, and the ground state becomes a pure $d$-wave
superconductor.
All the described states have a gapless excitation spectrum -- note here
that the flux phase can be interpreted as a $d_{x^2-y^2}$ density wave,
implying nodes of the order parameter in the diagonal directions in momentum
space.

At very large values of $V$ the dominating repulsion gives rise
to Wigner crystal states as described above.
However, there are no pair crystal phases for the cases $x=0.5$ and $0.6$
which is related to the absence of phase separation tendencies
at these parameter values.

Finally we note that the case $x=1$, i.e., the ``pure'' SU($N$) interaction
decoupling, leads to phase separation in the absence of Coulomb repulsion $V$ \cite{grilli},
and the inclusion of $V$ is expected to produce ``striped'' ground
states \cite{sunstripes}.

\subsection{Phase transitions at $N=\infty$}
\label{sec:phasetran}

We briefly discuss the nature of the transitions between the
various ground state phases at the mean-field level
($N=\infty$).

The transition from a $d$-wave superconductor, with ${\cal C}$ unbroken,
to the fully-formed $p \times 1$ stripes discussed above, can
either be first-order, or via intermediate states with partial
stripe order.
In the latter case, there is first a continuous transition to a
doped spin-Peierls state with ${\cal C}$ symmetry breaking at $p=2$.
To our knowledge
a $p=2$ charge-ordered superconducting state has not been
experimentally detected, but a search for one should be
worthwhile. Then there is a second second-order transition to
$p=4$ state with partial stripe order, before the fully-formed
$p=4$, $q=2$ state with intervening insulating stripes appears.
In any case, the excitation spectrum in the stripe states is fully
gapped.
The transitions between the different stripe phases where $p$
changes its value are in general first-order since they arise
from level crossings.

At very small hole densities the Cooper pair crystal states are favored
over striped states, then the ground state nature changes from
one-dimensional to two-dimensional $\cal C$ breaking.
Again, this transition is either first-order or second-order
depending on the values of $t$, $J$, and $V$, in the latter case it
can be visualized as additional $\cal C$ breaking in the
longitudinal stripe direction.

Finally, the transition to a Wigner crystal phase of single charges
at large $V$ is always a first-order transition because the
inherent tendency to fermion pairing is broken in the Wigner
crystal.

By adding a biquadratic exchange term to the Hamiltonian we
found a $\cal T$-breaking superconductor with $s^{*}+id$ symmetry
and a gapped spectrum. As displayed in the phase diagram of
Fig.~\ref{figmf4}, it shows a second-order transition to
a pure $d$-wave state with increasing doping.
Similarly, the introduction of a second-nearest neighbor exchange
$J'$ can lead to a $\cal T$-breaking $(d_{x^2 - y^2}+id_{xy})$-wave
superconductor which also has a second-order transition to
a pure $d$-wave state with increasing $\delta$, see
Fig.~\ref{figmf4b}.

The SU(2$N$)-like interaction term $\bar{\cal H}_J$ (Sec. \ref{mixint})
which has to be decoupled in the particle-hole channel leads to other
possible transition scenarios.
As shown in Fig.~\ref{figmf5}, a $\cal T$-conserving spin-Peierls phase is
possible.
Then two transitions are found upon decreasing doping:
First there is a second-order transition from the $d$-wave state to a
(weakly dimerized) spin-Peierls state where
the dimerization sets in. The four nodes of the $d$-wave phase are
preserved, but shift gradually in momentum space.
At smaller $\delta$ and a ``critical'' dimerization there is another
second-order transition where the spectrum becomes gapped.
Turning to the $x=0.6$ phase diagram of Fig.~\ref{figmf6}, we note that
the transition between the OAF+$d$-wave coexistence phase and the
pure $d$-wave state is second-order.
Both order parameters have $d_{x^2-y^2}$ symmetry leading to gapless
spectra.
The $\cal T$ symmetry is broken in the low doping phase with orbital
currents.


\section{Continuous quantum phase transitions}
\label{sec:qpt}

In Section~\ref{sec:phasetran} we noted a number of continuous
quantum phase transitions in the mean-field phase diagrams of
Fig.~\ref{figmf1}-\ref{figmf6}. As we discussed in
Section~\ref{sec:qpt1}, for the purposes of understanding the
photoemission experiments of Ref~\onlinecite{valla}, we are
particularly interested in transitions in class A. Of these, the
simplest involve ${\cal T}$ symmetry breaking alone, and we will
consider these in the first subsection below. Transitions
involving breaking of ${\cal C}$
symmetry alone are considered next in Section~\ref{sec:cd}. This is
followed by a
discussion of the onset of staggered flux order
in a $d$-wave superconductor (this transition breaks
${\cal T}$ and ${\cal C}$ symmetries)
in Section~\ref{sec:oafd}.

\subsection{$\lowercase{d}$-wave to
$(\lowercase{s}^{*}+i\lowercase{d})$-wave or
$(\lowercase{d}_{x^2-y^2}+i\lowercase{d}_{xy})$-wave }
\label{sec:dsid}

A continuous transition between a $d$-wave superconductor and a
$(s^{*}+id)$-wave superconductor appears in the mean-field phase diagram
in Fig.~\ref{figmf4}, and fluctuations in its vicinity will be
described below. A very closely related theory applies to the
transition between a $d$-wave superconductor and a
$(d_{x^2-y^2} + i d_{xy})$-wave superconductor shown in Fig~\ref{figmf4b}.
Indeed,
to the order we are computing things, the results for the
scaling functions of the two transitions are identical.

We note that a transition between $d$ and $(d_{x^2-y^2} + i d_{xy})$
pairing was also considered recently by D.-H.~Lee \cite{dunghai}
using a rather different framework. Lee concluded that the
transition was in the universality class of the Ising model in a
transverse field; this is equivalent to assuming that the critical
theory is described by $S_{\phi}$ in (\ref{dsid3}) below, and
places the transition in class B. In fact, as we show below,
it is also necessary to include fermionic excitations: the
complete theory is in (\ref{dsid7}) below, and the transition is in
a new universality class which belongs in class A.
In a separate recent work, Balatsky {\em et al.} \cite{clap}
considered a ``clapping'' collective mode in a $(d_{x^2-y^2} + i d_{xy})$
superconductor: this is an
`angular' fluctuation mode expected in
a state with well-formed $d_{xy}$ pairing induced by an external magnetic
field; in contrast we are interested here in {\em spontaneous}
fluctuations to  $(d_{x^2-y^2} + i d_{xy})$ order, which will
dominated by `amplitude' fluctuations represented by our field
$\phi$ below.

We begin by establishing notation, and reviewing the familiar physics of the low-energy
fermionic quasiparticles in a $d$-wave superconductor. For the most part, we
will follow the notation of Balents {\em et al.}~\cite{balents}. The
low energy excitations appear at four points in the Brillouin
zone: $(K,K)$, $(-K,K)$, $(-K,-K)$, and $(K, -K)$. We denote the
components of the electron annihilation operator, $c_a$, in their vicinity
by $f_{1a}$, $f_{2a}$, $f_{3a}$, $f_{4a}$ respectively,
where $a= \uparrow, \downarrow$ is the electron spin component.
Now introduce the 4-component Nambu spinors $\Psi_1 =
(f_{1a}, \varepsilon_{ab} f_{3b}^{\dagger})$
and  $\Psi_2 =
(f_{2a}, \varepsilon_{ab} f_{4b}^{\dagger})$ where
$\varepsilon_{ab}$ is an antisymmetric tensor with
$\varepsilon_{\uparrow \downarrow} = 1$.
The action for the fermionic excitations in the $d$-wave
superconductor is then
\begin{eqnarray}
S_{\Psi} &=& \int \!\!\! \frac{d^d k}{(2 \pi)^d} T \!\! \sum_{\omega_n}
\Psi_1^{\dagger}  \left(
- i \omega_n + v_F k_x \tau^z + v_{\Delta} k_y \tau^x \right) \Psi_1  \nonumber \\
&~& \!\!\!\!\!\!\!\!\!\!\!\!\!\!\!\!\!\! +\int \!\!\! \frac{d^d k}{(2 \pi)^d} T \!\! \sum_{\omega_n}
\Psi_2^{\dagger}  \left(
- i \omega_n + v_F k_y \tau^z + v_{\Delta} k_x \tau^x \right) \Psi_2 .
\label{dsid1}
\end{eqnarray}
Here $\tau^{\alpha}$ are Pauli matrices which act in the fermionic
particle-hole space, $k_{x,y}$ measure the wavevector from the nodal points and
have been rotated
by 45 degrees from the axes of the square lattice, and $v_{F}$, $v_{\Delta}$
are velocities.

Let us now consider the transition to the $(s^{*}+id)$-wave
superconductor. This involves only a $Z_2$ symmetry breaking of
the ${\cal T}$ symmetry, and so the order parameter is a real
scalar field, $\phi$. In the presence of a non-zero,
spacetime-independent $\phi$, the superconducting gap function
takes the form
\begin{equation}
\langle c_{k \uparrow} c_{-k \downarrow} \rangle
= \Delta_0 (\cos k_x - \cos k_y) + i \phi (\cos k_x + \cos k_y).
\label{dsid2}
\end{equation}
The superconducting order parameter also has a single overall
complex phase, but because the superfluid stiffness is finite, its
fluctuations can be neglected in the critical theory.
On general symmetry grounds, we can write down the following
effective action for $\phi$ fluctuations
\begin{equation}
S_{\phi} = \int \!\! d^d x d \tau \Big[
\frac{1}{2}(\partial_{\tau} \phi)^2 + \frac{c^2}{2} (\nabla \phi )^2 +
\frac{s_0}{2} \phi^2 + \frac{u_0}{24} \phi^4 \Big];
\label{dsid3}
\end{equation}
Here $c$ is a velocity, $s_0$ is the parameter which tunes the
system across the transition, and $u_0$ measures the quartic
self-interaction of the order parameter.
Normally, in the absence of strict particle-hole symmetry, a term
with a first-order time derivative can potentially appear in $S_{\phi}$;
however, $\phi$ is a real field, and the only possible relevant term,
$\phi \partial_{\tau} \phi = (1/2) \partial_{\tau} \phi^2$,
is a total derivative and integrates to zero.

The final piece of the action is the coupling between $\phi$ and
the fermionic excitations. This can be easily deduced from
(\ref{dsid2}), and the standard pairing interaction between
$\Delta(k)$ and the electrons. In the vicinity of the nodal
points, this coupling reduces to the very simple expression
\cite{senthil} (we are assuming here that $K \neq \pi/2$, as is
the case experimentally \cite{valla}, so that $(\cos k_x + \cos k_y)$
does not vanish at the nodal points)
\begin{equation}
S_{\Psi\phi} = \int \!\! d^d x d \tau \Big[ \lambda_0 \phi
\left( \Psi_1^{\dagger} \tau^y \Psi_1 + \Psi_2^{\dagger} \tau^y
\Psi_2 \right) \Big].
\label{dsid4}
\end{equation}
The remainder of this subsection will consider the properties
of the theory $S_{\Psi} + S_{\phi} + S_{\Phi\phi}$. Before we
embark on this case, we note the generalization
to the case of a transition to $d_{x^2-y^2} + i d_{xy}$ order: in
this case (\ref{dsid2}) is replaced by
\begin{equation}
\langle c_{k \uparrow} c_{-k \downarrow} \rangle
= \Delta_0 (\cos k_x - \cos k_y) + i \phi \sin k_x \sin
k_y ;
\label{dsid5}
\end{equation}
in the vicinity of the nodal points, instead of (\ref{dsid4}), we
obtain \cite{senthil}
\begin{equation}
\widetilde{S}_{\Psi\phi} = \int \!\! d^d x d \tau \Big[ \lambda_0 \phi
\left( \Psi_1^{\dagger} \tau^y \Psi_1 - \Psi_2^{\dagger} \tau^y
\Psi_2 \right) \Big].
\label{dsid6}
\end{equation}
The only change is the relative minus sign between the two terms, which arises
from the changing sign of the factor $\sin k_x \sin k_y$ in (\ref{dsid5}) between the
nodal points;
it is this change which is responsible for the non-zero spin Hall
conductance of the $d_{x^2-y^2} + i d_{xy}$ state \cite{senthil}.
The properties of the theory $S_{\Psi} + S_{\phi} + \widetilde{S}_{\Psi\phi}$
are very closely related to that of $S_{\Psi} + S_{\phi} +
S_{\Psi\phi}$; indeed, at the one-loop order we compute things
here, there is no difference between the theories, and our results
can be applied equally to the transition to either $s^{*}+i d$
or $d_{x^2-y^2} + i d_{xy}$ order.

Our renormalization group analysis is closely related to that
discussed by Zinn-Justin \cite{zj}.
We will mainly restrict our study to the case in which all the
velocities in $S_{\Psi,\phi}$ are equal to each other.
By a suitable choice of the scale of time we can then set
$c=v_F=v_{\Delta}=1$ at the outset. Deviations from exactly equal
velocities will be considered in Section~\ref{sec:anisotropy}
below, and in future work. With exactly equal velocities, the
action is actually Lorentz invariant and so we must have
the dynamic exponent $z=1$; this will value of $z$ will be
implicitly assumed in the remainder of this paper.
We introduce notation to
make the Lorentz invariance explicit.
We map $\Psi_2 \rightarrow (\tau^x + \tau^z) \Psi_2 /\sqrt{2}$,
and define $\overline{\Psi}_{1,2} = - i \tau^y \Psi_{1,2}^{\dagger}
$, and introduce Lorentz index $\mu = \tau,x,y$.
Then $S_{\Psi} + S_{\phi} + S_{\Psi\phi}$
can be written in the compact Lorentz-invariant form
\begin{eqnarray}
S_{s^{*}+id} &=& \int d^D x \Big(
i \overline{\Psi}_1 \partial_{\mu} \gamma^{\mu} \Psi_1 +
i \overline{\Psi}_2 \partial_{\mu} \gamma^{\mu} \Psi_2 +
\frac{1}{2} ( \partial_{\mu} \phi)^2 \nonumber \\
&~& \!\!\!\!\!\!\!\!\!\!\!\!
+ \frac{s_0}{2} \phi^2 + \frac{u_0}{24} \phi^4
- i \lambda_0 \phi ( \overline{\Psi}_1 \Psi_1 - \overline{\Psi}_2
\Psi_2) \Big),
\label{dsid7}
\end{eqnarray}
where $D=d+1$, and $\gamma^{\mu} = (-\tau^y, \tau^x, \tau^z)$.
We note that there is also a corresponding action
$S_{d_{x^2-y^2}+id_{xy}}$ which (using (\ref{dsid6})) differs only
in having a relative plus sign between the couplings of $\overline{\Psi}_1
\Psi_1$ and $\overline{\Psi}_2 \Psi_2$ to $\phi$.

We now proceed with the renormalization group analysis of
$S_{s^{*}+id}$ or $S_{d_{x^2-y^2}+id_{xy}}$.
We will present the results using the field-theoretic
method as it leads to a more streamlined discussion of finite
temperature properties. To this end, we introduce the
wavefunction renormalizations
\begin{eqnarray}
\phi &=& Z_b^{1/2} \phi_{R} \nonumber \\
\Psi_{1,2} &=& Z_f^{1/2} \Psi_{1,2R}
\label{dsid9}
\end{eqnarray}
and the coupling constant renormalizations
\begin{eqnarray}
\lambda_0 &=& \frac{\mu^{\epsilon/2}}{S_{D}^{1/2}}
\frac{Z_{\lambda}}{Z_f Z_b^{1/2}}\  \lambda \nonumber \\
u_0 &=& \frac{\mu^{\epsilon}}{S_{D}}
\frac{Z_{u}}{Z_b^{2}}\  u
\label{dsid10}
\end{eqnarray}
where $\mu$ is a renormalization momentum scale and $S_D =
2/(\Gamma(D/2) (4 \pi)^{D/2})$.

One loop calculations of the renormalization constants using
minimal subtraction of poles in $\epsilon=4-D=3-d$ yield
\begin{eqnarray}
Z_b &=& 1 - \frac{4 \lambda^2}{\epsilon} \nonumber \\
Z_f &=& 1 - \frac{\lambda^2}{2\epsilon} \nonumber \\
Z_{\lambda} &=& 1 + \frac{\lambda^2}{\epsilon} \nonumber \\
Z_u &=& 1 + \frac{3 u}{2 \epsilon} - \frac{48 \lambda^4}{u
\epsilon}.
\label{dsid11}
\end{eqnarray} From these we obtain the beta-functions (note that we are using
the field-theorists' definition of the beta function \cite{zj}, and these
are opposite in sign to those normally used by condensed matter physicists):
\begin{eqnarray}
\beta(\lambda) &=& -\frac{\epsilon}{2} \lambda + \frac{7}{2}\lambda^3
\nonumber \\
\beta(u) &=& - \epsilon u + \frac{3}{2} u^2  + 8 u \lambda^2  - 48
\lambda^4.
\label{dsid12}
\end{eqnarray}
These equations have the infrared stable fixed point
\begin{eqnarray}
\lambda^{\ast 2} &=& \frac{\epsilon}{7} \nonumber \\
u^{\ast} &=& \frac{16\epsilon}{21}
\label{dsid13}
\end{eqnarray}
which controls the physics in the vicinity of the quantum critical
point. In particular, the fields acquire the anomalous dimensions
\begin{eqnarray}
\eta_b &=& 4 \lambda^{\ast 2} \nonumber\\
\eta_f &=& \lambda^{\ast 2}/2
\label{dsid14}
\end{eqnarray}
which will play an important role in the spectral functions to be
discussed below.

In the first subsection below, we will consider the consequences
of unequal velocities.
The next subsection will describe the $T>0$ quantum critical
spectral response functions of the fixed point (\ref{dsid14}).

\subsubsection{Unequal velocities}
\label{sec:anisotropy}

The full treatment of the problem with unequal velocities is
considerably more complicated than the simple renormalization
group analysis above, and is deferred to future work.
It is entirely possible that there are other non-Lorentz-invariant
fixed points describing the critical theory: however, apart from
differences in the details of the scaling functions, the
qualitative properties of such fixed points should be quite
similar to the Lorentz-invariant case explored in more detail in
Section~\ref{sec:damp}.

We will restrict our attention here to perturbations which break
Lorentz invariance in a linear stability analysis
of the Lorentz-invariant fixed point. Such an analysis was also carried out in
Ref.~\onlinecite{balents}, but the authors do not appear to have
performed the proper decomposition of the perturbations into the appropriate
eigenoperators, as is required to obtain the correct scaling
dimensions.

It is important to classify the perturbations in terms of
irreducible representations of the Lorentz symmetry of the fixed
point. For the velocity differences, these are the
symmetric, traceless, second-rank Lorentz tensors. So we consider
the following perturbation to the action
$S_{\Psi}+S_{\phi}+S_{\Psi\phi}$:
\begin{eqnarray}
&& S_a = \int d^D x \Biggl[
\widetilde{g}_{\mu\nu} \left(\partial_{\mu} \phi \partial_{\nu} \phi -
\frac{\delta_{\mu\nu}}{D} ( \partial_\rho \phi)^2 \right)
\nonumber \\
&& \!\!\!\! + i g^{(1)}_{\mu \nu} \overline{\Psi}_1 \left(
\frac{\partial_{\mu} \gamma^{\nu} + \partial_{\nu} \gamma^{\mu}}{2} -
\frac{ \delta_{\mu \nu}}{D} \partial_{\rho} \gamma^{\rho} \right)
\Psi_1 \nonumber \\
&& \!\!\!\! + i g^{(2)}_{\mu \nu} \overline{\Psi}_2 \left(
\frac{\partial_{\mu} \gamma^{\nu} + \partial_{\nu} \gamma^{\mu}}{2} -
\frac{\delta_{\mu \nu}}{D} \partial_{\rho} \gamma^{\rho} \right)
\Psi_2
  \Biggr] .
\label{dsid8}
\end{eqnarray}
It is crucial that the tensors above are traceless: this ensures,
at linear order,
that such perturbations do not mix with terms already in the
Lorentz-invariant action, and with redundant operators which can
be removed by a rescaling of the fields which are being integrated
over.

It is a simple matter to compute the one-loop renormalization
group flow equations to the terms in (\ref{dsid8}). We find
\begin{equation}
\left( \begin{array}{c} \beta(\widetilde{g}_{\mu\nu} ) \\
\beta( g^{(1)}_{\mu\nu} ) \\
\beta( g^{(2)}_{\mu\nu} ) \end{array} \right)
= \lambda^2 \left( \begin{array}{ccc}
4 & -2 & -2 \\
-1/3 & 1/3 & 0 \\
-1/3 & 0 & 1/3 \end{array} \right)
\left( \begin{array}{c} \widetilde{g}_{\mu\nu}  \\
g^{(1)}_{\mu\nu}  \\
g^{(2)}_{\mu\nu} \end{array} \right).
\label{dsid15}
\end{equation}
Clearly, we need the eigenvalues and eigenvectors of the matrix of
coefficients in (\ref{dsid15}). First, note that there is an
eigenvector, $(1,1,1)$, with zero eigenvalue. This is expected, and its
presence is an important check on our calculation---it
corresponds to the perturbation of simply changing {\em
all} velocities by the same amount: such a redundant perturbation can be
absorbed by a rescaling of spacetime co-ordinates, and does not
modify any of the physics.

The other eigenvalues and eigenvectors correspond to physical
differences in the velocities. The eigenvector $(0,1,-1)$ has
eigenvalue $1/3$ and corresponds to the perturbation in which
$v_F$ and $v_{\Delta}$ are made unequal. So such a perturbation is
irrelevant in the infrared, with scaling dimension $-\lambda^{\ast 2}/3 =
-\epsilon/21$. The final eigenvector, $(-12,1,1)$, induces a
difference between the velocities of the fermions and bosons, and
this perturbation is more strongly irrelevant: the scaling
dimension is $-13 \lambda^{\ast 2}/3 =
-13\epsilon/21$.

So we have established that the Lorentz-invariant fixed point is
at least linearly stable to velocity differences. However, the
scaling dimension of the leading irrelevant operator is quite
small $(-\epsilon/21)$, and this could lead to very slowly decaying
transients in the
critical behavior.

\subsubsection{$T>0$ spectral functions}
\label{sec:damp}

The problem of computing $T>0$, quantum critical, spectral
functions in 2+1 dimensions is one of considerable complexity.
Even though the non-linear couplings at the fixed point,
(\ref{dsid3}), are small for small $\epsilon$, the low frequency
response at $T>0$ cannot be computed in a bare $\epsilon$ expansion.
This is because the limits $\epsilon \rightarrow 0$
and $\omega \rightarrow 0$ do not commute, and there are
spurious low-frequency singularities in the bare $\epsilon$
expansion: this is discussed further in Appendix~\ref{app:damp}.
For the case of a scalar field alone, this problem of $T>0$
dynamics was addressed in Refs.~\onlinecite{ssepsilon,book}
by a two-step procedure involving a mapping to a quasi-classical
problem. Here we shall use the insight gained from these previous,
controlled studies to motivate a simple self-consistent one-loop
theory for computing the low frequency relaxation rates. In
contrast to
the previous studies \cite{ssepsilon}, we are aided here by the
fact that damping appears already at the one-loop level in
$S_{s^{*}+id}$ or $S_{d_{x^2-y^2}+id_{xy}}$, and this simplifies the solution of the
self-consistent equations.

We shall be interested in the following retarded response functions
in the ``quantum critical'' \cite{book} regime:
\begin{eqnarray}
G_f(k, i\omega_n) &=& Z_f^{-1} \left\langle \Psi_1 (k, \omega_n)
\Psi_1^{\dagger} (k, \omega_n )\right\rangle \nonumber \\
G_b(k, i\omega_n) &=& Z_b^{-1} \left\langle \phi (k, \omega_n)
\phi(-k, -\omega_n ) \right\rangle.
\label{dsid16}
\end{eqnarray}
After analytic continuation to real frequencies
$G_f$ satisfies the scaling form in
(\ref{i1}), and an analogous result for $G_b$ with $f \rightarrow
b$.

For $\omega \gg T$ or $k \gg T$, the response functions
in (\ref{dsid16}) equal their $T=0$ results, and these can be
computed in the bare $\epsilon$ expansion. A familiar computation
in the one loop renormalized theory gives us
\begin{eqnarray}
G_{f} (k, \omega) &=& C_f \mu^{-\eta_f}  \frac{ \omega + k_x \tau^z + k_y
\tau^x}{\left[ k^2 - (\omega+ i0)^2 \right]^{1-\eta_f/2}}
\nonumber \\
G_{b} (k, \omega) &=& C_b \mu^{-\eta_b} \frac{1}{\left[ k^2
- (\omega+ i0)^2 \right]^{1-\eta_b/2}}
\label{dsid17}
\end{eqnarray}
where $C_{f,b}$ are constants given by
\begin{eqnarray}
C_f &=& 1 - \lambda^{\ast 2}/4 + \ldots \nonumber \\
C_b &=& 1 - 2 \lambda^{\ast 2} + \ldots.
\label{dsid18}
\end{eqnarray}
Clearly, (\ref{dsid17}) obeys the scaling form (\ref{i1}).

Next we consider the more difficult case where $\omega, k$ are of order $T$
or smaller. Here we shall proceed by using an ansatz for the low
frequency form of the Green's functions: this ansatz is motivated
by the results of the corresponding low frequency regime in a
number of other models \cite{ssepsilon,book}, especially the
exactly solvable Ising chain in a transverse field \cite{book}. In
these studies, it was found that, provided the number of order
parameter components was not large, the low frequency spectral
functions had a simple relaxational form, that of degrees of
freedom in a simple dissipative medium. In the present situation,
we have a single scalar order parameter, and can expect that a
similar situation should apply. We therefore make the ansatz
\begin{eqnarray}
G_{f} (k, \omega) &=& \left( \frac{\mu}{T} \right)^{-\eta_f}
\frac{\omega + i \Gamma_f + k_x \tau^z + k_y \tau^x}{k^2 - (\omega
+ i \Gamma_f)^2 } \nonumber \\
G_{b} (k, \omega) &=&  \left( \frac{\mu}{T} \right)^{-\eta_b} \frac{1}{k^2 -
(\omega +  i  \Gamma_b )^2 }.
\label{dampr}
\end{eqnarray}
where $\Gamma_{f,b}$ are damping frequencies we need to determine.
The functional form for $G_b$ in (\ref{dampr}) was found to be a
remarkably good fit to the low frequency portion of the known exact
result for the spectral function of an Ising chain in a transverse
field \cite{book,note}. A further rationale
behind (\ref{dampr}) is that we have only
included terms in the self energy of the fields which are lower powers of the
frequency than those already present in the free propagator, and
which are consistency with the requirement that all spectral
functions are smooth at $\omega=0$ for $T>0$. It is precisely
these low frequency powers which acquire a singular form in the
bare $\epsilon$ expansion, and so we are forced to use the
self-consistent approach to be described below. Comparing
(\ref{dampr}) with (\ref{i1}) we see that consistency demands that
$\Gamma_{f,b}$  be universal numbers times temperature.
The purpose of the remainder of this subsection is to determine
the universal numbers $\Gamma_f/T$ and $\Gamma_b/T$.
Before turning to this, let us also quote, the form
of (\ref{dampr}) in imaginary frequencies
\begin{eqnarray}
G_{f} (k, i\omega_n) &=&   \left( \frac{\mu}{T} \right)^{-\eta_f} \frac{i \omega_n + i
\Gamma_f \mbox{sgn}(\omega_n) + k_x \tau^z + k_y \tau^x}{k^2 +
(|\omega_n | +  \Gamma_f)^2} \nonumber \\
G_{b} (k, i \omega_n) &=&   \left( \frac{\mu}{T} \right)^{-\eta_b}
\frac{1}{k^2 + (|\omega_n| +
\Gamma_b)^2 }
\label{damp}
\end{eqnarray}

The damping represented by $\Gamma_{f,b}$ arises from interactions
between the thermally excited bosons and fermions. The typical
excitation will have energy of order $T$, and as the damping is
dominated by the lowest energy excitations, the typical
interaction vertex will have external frequencies at order $T$ or
lower. Motivated by this, we develop a perturbation theory for the
interactions not in terms of the bare vertices, but in terms of the
full interaction vertices between the
excitations. At the one-loop level, the damping terms arise from
the $\lambda_0$ interactions alone: however rather than expanding
in powers of $\lambda_0$, we will express the self energy in
powers of the full three-point irreducible vertex, $\Lambda_3$
between one boson ($\phi$) and two Fermi ($\Psi_{1,2}^{\dagger}$,
$\Psi_{1,2}$) fields. A convenient choice is to use the zero
momentum vertex between a Bose field at zero frequency, and Fermi
fields at the minimum Matsubara frequency of $\epsilon_n = \pi T$.
Bare perturbation theory for this vertex gives
\begin{equation}
\Lambda_3 = \lambda_0 - \lambda_0^3 \int \frac{d^d k}{(2\pi)^d}
T \sum_{\omega_n} \frac{1}{(k^2+ \omega_n^2)(k^2+ (\epsilon_n +
\omega_n)^2)}
\label{d1}
\end{equation}
where $\omega_n$ is a fermionic frequency.
General scaling arguments from renormalization theory \cite{ssepsilon,bgz}
show that at
the fixed point the exact result for the vertex obeys
\begin{equation}
\Lambda_3 = \frac{T^{\epsilon/2}}{Z_f Z_b^{1/2}}
\left( \frac{\mu}{T} \right)^{\eta_f+\eta_b/2} C_3
\label{d2}
\end{equation}
where $C_3$ is a {\em universal} number.
Evaluating the
frequency sums and the momentum integrals in (\ref{d1}), and expressing
everything in terms of renormalized couplings at the fixed point,
and collecting low order terms in $\epsilon$, we find that
(\ref{d2}) is indeed obeyed, with the universal number
\begin{equation}
C_3 = \frac{1}{S_{d+1}^{1/2}} \left( \lambda^{\ast} + 0.374367
\lambda^{\ast 3} + \ldots \right)
\label{d2a}
\end{equation}

To proceed with the damping calculation, we write down the
structure of the self-consistent one-loop equations in terms
of the Green's functions of the bare fields; from (\ref{dsid16})
these are $G_f^B = Z_f G_f$ and $G_b^B = Z_b G_b$.
 We define the self
energies by
\begin{eqnarray}
\left( G_f^{B} \right)^{-1} &=& -i \omega_n + k_x \tau^z + k_y \tau^x -
\Sigma_f^B
\nonumber\\
\left( G_b^{B} \right)^{-1} &=& \omega_n^2 + k^2 + s_0 - \Sigma_b^B.
\label{defsig}
\end{eqnarray}
The self-consistent, one-loop expression for the self energies, expressed in terms
of $\Lambda_3$, are
\begin{eqnarray}
\Sigma_f^B (k, \omega_n) &=&  \Lambda_3^2 \int \!\!\frac{d^d p}{(2
\pi)^d} T \sum_{\epsilon_n} \tau^y G_f^B (p,\epsilon_n) \tau^y
\nonumber \\
&~&~~~~~~~~~\times G_b^B (k-p,\omega_n-\epsilon_n) \nonumber \\
\Sigma_b^B (k, \omega_n) &=& - 2 \Lambda_3^2 \mbox{Tr} \int \!\!\frac{d^d p}{(2
\pi)^d} T \sum_{\epsilon_n} \tau^y G_f^B (p,\epsilon_n) \tau^y
\nonumber \\
&~&~~~~~~~~~\times G_f^B (k+p,\omega_n+\epsilon_n)
\label{d3}
\end{eqnarray}
We will now express this in terms of renormalized quantities.
First we note by comparing (\ref{defsig}) with (\ref{dampr},\ref{damp}) that
we only need the imaginary part of the self energies at
small real frequencies; in particular, the damping co-efficients
can be expressed as
\begin{eqnarray}
\Gamma_f &=&  \left(\frac{\mu}{T}\right)^{-\eta_f}  Z_f \lim_{\omega
\rightarrow 0} \mbox{Im} \Sigma_f^B (0,\omega) \nonumber \\
\Gamma_b  &=& \left(\frac{\mu}{T}\right)^{-\eta_b} Z_b \lim_{\omega
\rightarrow 0} \frac{\mbox{Im} \Sigma_b^B (0,\omega)}{2 \omega}
\label{d4}
\end{eqnarray}
We now insert (\ref{damp}), (\ref{d2}), (\ref{d3}) into
(\ref{d4}). To our satisfaction, we find that all factors of the
renormalization factors $Z_{b,f}$ and the scale $\mu$ precisely
cancel out, and the remaining expressions involve only universal
quantities.
Performing the frequency summation in (\ref{d3})  and inserting
in (\ref{d4}) we obtain
\begin{eqnarray}
\Gamma_f &=&  C_3^2  \int_0^{\infty} \! \frac{d
\Omega}{\pi} \int \!\! \frac{d^d k}{(2 \pi)^d}
\frac{ T^{\epsilon}}{\sinh(\Omega/T)}
\nonumber \\
&~& \times
\mbox{Im} \left( \frac{1}{k^2 - (\Omega + i \Gamma_b )^2} \right) \nonumber \\
&~& \times
\mbox{Im} \left( \frac{\Omega + i \Gamma_f}{k^2 - (\Omega + i
\Gamma_f)^2} \right) \nonumber \\
\Gamma_b  &=& 2 C_3^2
\int_0^{\infty} \! \frac{d
\Omega}{\pi} \int \!\! \frac{d^d k}{(2 \pi)^d}
\frac{T^{\epsilon-1}}{\cosh^2(\Omega/2T)}
\nonumber \\
&~& \times \Bigg[\left\{\mbox{Im} \left( \frac{\Omega + i \Gamma_f}{k^2 - (\Omega + i
\Gamma_f)^2} \right)\right\}^2
\nonumber \\
&~& \!\!\!\!\!\!\!\!\!\!\!\!\!\!\!\!
- k^2 \left\{\mbox{Im} \left( \frac{1}{k^2 - (\Omega + i
\Gamma_f)^2} \right)\right\}^2 \Bigg]
\label{d5}
\end{eqnarray}
A simple dimensional analysis of (\ref{d5}) shows that $T$ can be
completely scaled out of both equations for all $d$. The strength of the
damping is determined by the dimensionless ratios $\Gamma_f/T$
and $\Gamma_b/T$, and these are completely determined by the
dimensionless universal $C_3$.

We solved (\ref{d5}) numerically in $d=2$.
The results are
$\Gamma_f/T=0.581$ and $\Gamma_b/T = 0.170$.

\subsection{${\cal C}$ symmetry breaking in a $\lowercase{d}$-wave superconductor}
\label{sec:cd}

A number of transitions involving ${\cal C}$ symmetry breaking
in a $d$-wave superconductor were noted in Section~\ref{sec:spn}.
These involve the onset of either stripe or spin-Peierls order,
and such transitions appear in Figures~\ref{figmf1}, \ref{figmf3},
and~\ref{figmf5}. In all cases, the order parameter can be
identified with scalars $\Phi_x$, $\Phi_y$ representing the
amplitude of charge density waves with wavevectors $(Q,0)$ and
$(0,Q)$. If $Q$ is commensurate with the underlying lattice, then
$\Phi_{x,y}$ are real; otherwise $\Phi_{x,y}$ are complex, with
their phases representing the freedom of the charge density wave
to slide with respect to the underlying lattice.
On general symmetry grounds, we can write down an effective action
for $\Phi_{x,y}$, similar to (\ref{dsid3}):
\begin{eqnarray}
 S_{\Phi} &=& \int d^d x d \tau \Big[
|\partial_{\tau} \Phi_x|^2 + |\partial_{\tau} \Phi_y |^2 +
|\nabla \Phi_x|^2 + |\nabla \Phi_y|^2 \nonumber \\
&~&~~~~+ s_0 \left( |\Phi_x|^2 + |\Phi_y|^2 \right)
+ \frac{u_0}{2}
\left( |\Phi_x|^4 + |\Phi_y|^4 \right) \nonumber \\
&~&~~~~ + v_0 |\Phi_x|^2 |\Phi_y|^2
\Big],
\label{cd1}
\end{eqnarray}
where, for now, $\Phi_{x,y}$ can be either real or complex.
First-order time derivative terms, like $\Phi_x^{\ast} \partial_{\tau}
\Phi_x$,
are forbidden here by spatial inversion symmetry under
which $\Phi_x \rightarrow \Phi_x^{\ast}$, and such a term
changes sign after integration by parts.

To complete the theory, we have to consider the coupling of
$\Phi_{x,y}$ to the gapless Fermi excitations at wavevectors $(\pm
K, \pm K)$. Conservation of momentum implies that there is in fact
no long-wavelength coupling between $\Phi_{x,y}$ and $\Psi_{1,2}$
(which is linear in $\Phi_{x,y}$) unless $Q=2K$.
The mean-field studies of Section~\ref{sec:spn} always obtained $Q \neq 2K$,
and this expected to be the generic behavior. However, we cannot
rule out the possibility that there is a mode-locking phenomenon
which preferentially condenses a charge density wave at wavevector
$Q=2K$ over a finite range of parameters.

For $Q\neq 2K$,
$S_{\Phi}$ is the complete critical theory of the transition: the
fermions are not part of the critical theory and so the transition
is in class B. The simplest allowed couplings between the fermions
and the critical degrees of freedom are terms like $w_0 \int d^d x
d \tau |\Phi_x|^2 \Psi^{\dagger}_1 \tau^z \Psi_1$. Simple power
counting shows that $w_0$ has scaling dimension $1/\nu - d$, where
$\nu$ is the correlation length exponent of the transition
described by (\ref{cd1}). We expect that this $\nu$ is greater
than that of the $d+1=3$ dimensional XY model, which is $\approx
2/3$, and hence $w_0$ is irrelevant in $d=2$. By counting scaling
dimensions (or by an explicit perturbative computation) we can
deduce that the self energy of the nodal fermions
obeys $\mbox{Im} \Sigma_f^B \sim w_0^2 T^{2d+1-2/\nu}$, and so the
damping rate vanishes with a super-linear power of $T$ as $T
\rightarrow 0$, as expected for a class B transition.

In the remainder of this section, we consider the class A
transition with $Q=2K$, and both incommensurate (experimentally,
and in our mean-field theory, $K$ is incommensurate), so that $\Phi_{x,y}$ are
complex. Now a
coupling between $\Phi_{x,y}$ and $\Psi_{1,2}$ is possible.
Writing down all possible terms consistent with symmetries we
obtain \cite{stripeprl}
\begin{eqnarray}
S_{\Psi\Phi} &=& \int \!\! d^d x d \tau \bigg[
(\lambda_0 + \zeta_0) \left(
\Phi_x \Psi_2^{\dagger} \tau^z \Psi_1 + \Phi_y \varepsilon_{ab}
\Psi_{2a} \tau^x \Psi_{1b} \right) \nonumber\\
&-&(\lambda_0 - \zeta_0) \left(
\Phi_x \Psi_2^{\dagger} \tau^x \Psi_1 + \Phi_y \varepsilon_{ab}
\Psi_{2a} \tau^z \Psi_{1b} \right)\nonumber \\
&~&~~~~~~~~~~~~~~~~~ + \mbox{H.c.} \bigg]
\label{cd2}
\end{eqnarray}
The renormalization group analysis of $S_{\Psi} + S_{\Phi}
+ S_{\Psi\Phi}$ parallels that carried out in
Section~\ref{sec:dsid}, and so we will be brief.
The theory can be shown to be Lorentz invariant
for $c=v_F=v_{\Delta}$ and $\zeta_0=0$, and so also has $z=1$.
For this Lorentz-invariant
case, the renormalization constants, replacing those in
(\ref{dsid11}) are
\begin{eqnarray}
Z_b &=& 1 - \frac{2 \lambda^2}{\epsilon} \nonumber \\
Z_f &=& 1 - \frac{\lambda^2}{\epsilon} \nonumber \\
Z_{\lambda} &=& 1 - \frac{\lambda^2}{\epsilon} \nonumber \\
Z_u &=& 1 + \frac{5 u^2+v^2-2 \lambda^4}{u\epsilon} \nonumber \\
Z_v &=& 1 + \frac{2 v^2 + 4 u v - 4 \lambda^4}{v\epsilon}
\label{cd3}
\end{eqnarray}
(the coupling $v$ has been defined from $v_0$ following the
relationship between $u$ and $u_0$ in (\ref{dsid10})),
the beta functions, replacing those in (\ref{dsid12}) are
\begin{eqnarray}
\beta(\lambda) &=& -\frac{\epsilon}{2} \lambda + \lambda^3
\nonumber \\
\beta(u) &=& - \epsilon u + 5 u^2  +v^2 + 4 u \lambda^2  - 2
\lambda^4 \nonumber \\
\beta(v) &=& - \epsilon v + 2 v^2 +4 u v + 4 v \lambda^2 - 4
\lambda^4,
\label{cd4}
\end{eqnarray}
and the anomalous dimensions modifying those in (\ref{dsid14}) are
\begin{eqnarray}
\eta_b &=& 2 \lambda^{\ast 2} \nonumber\\
\eta_f &=& \lambda^{\ast 2}.
\label{cd5}
\end{eqnarray}
The computation of the $T>0$ spectral functions proceeds as
before, but with the following changes:
({\em i\/}) In (\ref{d1}) and (\ref{d2a}), the $\lambda^3$ terms
have the opposite sign; ({\em ii\/}) In (\ref{d3}), the integrand
in the expression for $\Sigma_f^B$ has a prefactor $2
\Lambda_3^2$, while that for $\Sigma_b^B$ has a prefactor
$-\Lambda_3^2$; ({\em iii\/}) In (\ref{d5}), the integrand
in the expression for $\Gamma_f$ has a prefactor $2
C_3^2$, while that for $\Gamma_b$ has a prefactor
$C_3^2$.
The numerical values of the damping coefficients are now
$\Gamma_f/T=1.35$ and $\Gamma_b/T = 0.395$.

\subsection{OAF order in a $\lowercase{d}$-wave superconductor}
\label{sec:oafd}
We discuss here the transition between a $d$-wave superconductor
and the phase with coexisting OAF (or staggered flux) and $d$-wave
superconducting order \cite{wang,nayak} : such a transition appears in
Fig.~\ref{figmf6} at $\delta \approx 0.12$. We will show that this
transition is in class B (in the notation of
Section~\ref{sec:qpt1}).

As has been emphasized by Nayak \cite{nayak}, the OAF is
characterized by a $d$-wave order parameter in the particle-hole
channel:
\begin{equation}
\langle c_{k+G,a}^{\dagger} c_{k,a} \rangle = i \phi
(\cos k_x - \cos k_y)
\label{oaf1}
\end{equation}
where $G=(\pi,\pi)$ and $\phi$ is a real order parameter. As in
the above subsections, the key issue is the coupling of this order
parameter to the fermionic quasiparticles of the $d$-wave
superconductor. As the order parameter carries momentum $G$, no
coupling, linear in $\phi$ is possible unless the nodal points are at
$(\pm K, \pm K)$ with $K=\pi/2$.

For $K \neq \pi/2$ the simplest
allowed coupling is $w_0 \int d^d x d \tau \phi^2 \Psi_1^{\dagger}
\tau^z \Psi_1 + \ldots$. As we saw in Section~\ref{sec:cd}, such a
coupling is irrelevant and places the transition in class B. The
damping rate obeys $\mbox{Im} \Sigma_f^B \sim w_0^2
T^{2d+1-2/\nu_I} \approx T^{1.83} $,
where $\nu_I \approx 0.63$ is the correlation length
exponent of the Ising model in $d+1=3$ (described by the field
theory $S_{\phi}$ in (\ref{dsid3})).

For completeness, let us also consider the special case where
$K=\pi/2$. Then, from (\ref{oaf1}) we can compute \cite{nayak} the following
coupling between $\phi$ and $\Psi_{1,2}$:
\begin{equation}
\phi \varepsilon_{ab} \Big[
i \Psi_{1a} \tau^x \partial_y \Psi_{1b} + i \Psi_{2a} \tau^x \partial_x \Psi_{2b}
+ \mbox{H.c.} \Big].
\label{oaf2}
\end{equation}
Note that this coupling has one more derivative than those in
(\ref{dsid6}) and (\ref{cd2}); this is a consequence of the vanishing
of the factor $(\cos k_x - \cos k_y)$ in (\ref{oaf1}) at the nodal points.
Therefore, (\ref{oaf2}) is irrelevant by
simple power-counting. A coupling such as (\ref{oaf2}) will not
lead to a fermion spectral function obeying (\ref{i1}); instead
the imaginary part of the fermionic self-energy vanishes as
$\mbox{Im} \Sigma_f^B \sim T^{2+\eta_I}$,
where $\eta_I>0$
is anomalous dimension of the order parameter ($\phi$) of the
Ising model in $D=3$ spacetime dimensions.
So even for $K=\pi/2$, this transition remains in
class B.



\section{Discussion}
\label{sec:conc}

This paper has presented a comprehensive mean-field study of
realistic models of the cuprate superconductors. A representative
sample of our results appears in Figs~\ref{figmf1}-\ref{figmf6},
and the properties of the phases therein were summarized in
Section~\ref{intro}. These mean-field results unify many other
earlier studies \cite{affmar,bza,gabi,biq,KotLiu,ekl,sr,grilli,ws,zeyer,sushkov}
and expose the relationships between them.

A second focus of the paper has been on the second-order quantum
phase transitions in Figs~\ref{figmf1}-\ref{figmf6}.
We paid particular attention to the $T>0$ fermionic quasiparticle
spectra in the vicinity of the nodal points in the Brillouin zone,
with the purpose of understanding the observed
quantum-criticality in recent photoemission experiments
\cite{valla}. We divided the transitions into two classes,
A and B (described in Section~\ref{sec:qpt1}), with only those in
class A leading to universal damping with $\omega/T$ scaling
near the nodal points.
Particularly appealing examples of class A transitions,
for which class A behavior was generic and did not require any
special parameter values, were those
involving time-reversal (${\cal T}$) symmetry breaking in a
$d$-wave superconductor: the most important of these are
the transitions from $d$-wave to
$(s^{*}+id)$-wave or $(d_{x^2-y^2} + i d_{xy})$-wave
superconductivity, which appeared in our
mean-field phase diagrams. The transition to $(d_{x^2 - y^2} + i
d_{xy})$ order had the additional satisfying feature of very
naturally leading to the absence of quantum-critical damping of quasiparticles
at momenta $(\pi,k)$, $(k, \pi)$ (with $0 \leq k \leq \pi$),
as is found in
experiments \cite{shen,campu,fed} below the superconducting critical temperature.
However, the
transition involving onset of ``staggered-flux'' (or orbital
antiferromagnet) order in a $d$-wave superconductor, which broke both
${\cal T}$ and ${\cal C}$ symmetries, was {\em not\/} of
class A. We note that fermion damping in a model involving
${\cal T}$ symmetry breaking has also been examined recently by
Varma \cite{cmv}, although he refers to quantum criticality associated with a
transition in a Fermi liquid and not a superconductor.
We also examined the onset of ``charge stripe-order'' (${\cal C}$ symmetry
breaking) in the $d$-wave
superconductor: such transitions belonged to class A if the charge
ordering wavevector was precisely equal to the separation between
two nodal points of the $d$-wave superconductor. This is a
fine-tuning condition, which is also not supported by experiments,
and makes the ${\cal C}$ breaking transition a less attractive
scenario for explaining fermion damping.

So the most viable candidate for the state $X$ in Fig~\ref{figx}
is the $(d_{x^2 - y^2} + i d_{xy})$-wave superconductor.  For this
case, the damping mechanisms appear to divide the
fermion excitations into two distinct components.
The fermions along the $(1,0)$, $(0,1)$ axes are strongly paired in the
$d_{x^2-y^2}$-wave state but are decoupled from the critical order parameter
fluctuations ($d_{xy}$) to the state $X$: consequently there is negligible
damping of these fermions below $T_c$. On the other hand, the
fermions along the $(1,\pm 1)$ axes couple strongly to the $d_{xy}$ order
parameter and undergo quantum critical damping as described by
(\ref{i1}). The situation changes dramatically once we go above
$T_c$. Now phase fluctuations and the proliferation of $hc/2e$
vortices \cite{phase} will strongly scatter
the fermions which couple efficiently to the predominant
$d_{x^2-y^2}$ order parameter: these are the fermions along
$(1,0), (0,1)$ directions, while the vortices are largely invisible
to the fermions along the $(1,\pm 1)$ directions. Moreover, the
antiferromagnetic spin fluctuations, which were responsible for
fermion pairing along the $(1,0), (0,1)$ axes below $T_c$, will
scatter these same fermions (on the ``hot spots'')
above $T_c$; again, these
fluctuations are invisible to the $(1, \pm 1)$ direction fermions
because the antiferromagnetic wavevector does not connect the
nodal points. Indeed, as we indicated in Section~\ref{sec:qpt1},
the predominant damping of the $(1, \pm 1)$ direction fermions
above $T_c$ continues to arise from the quantum critical $d_{xy}$
fluctuations to the state $X$: this mechanism applies as long as
the quantum-critical scattering length of these fermions remains
shorter than the superconducting phase coherence length.

With an eye towards comparisons with photoemission
experiments \cite{valla,mesot,mesot2},
we review our results for the nodal fermionic spectral functions of the
class A transitions. As in Ref.~\onlinecite{valla} we will follow
the evolution of the spectral function along a line from the
zone center going through the nodal points {\em e.g.} from (0,0)
along the (1,1) direction through the nodal point at $(K,K)$.
At the wavevector, $(K+k,K+k)$, our results for (\ref{i1}) are
contained in the diagonal components of (\ref{dsid17}, \ref{dampr});
we express these results in the form
\begin{equation}
G_f (k, \omega) = \left( \frac{\mu}{T} \right)^{-\eta_f}
\frac{1}{k - \omega - \widetilde{\Sigma}_f}.
\label{conc1}
\end{equation}
Note that $\widetilde{\Sigma}_f$ is strictly not a self-energy
(and thus the tilde), as
some of the self-energy corrections have already been absorbed
into the prefactor of an anomalous power of $T$ in (\ref{conc1}).
We have also set the velocity in the (1,1) direction to unity.

\begin{figure}
\epsfxsize=3.1in
\centerline{\epsffile{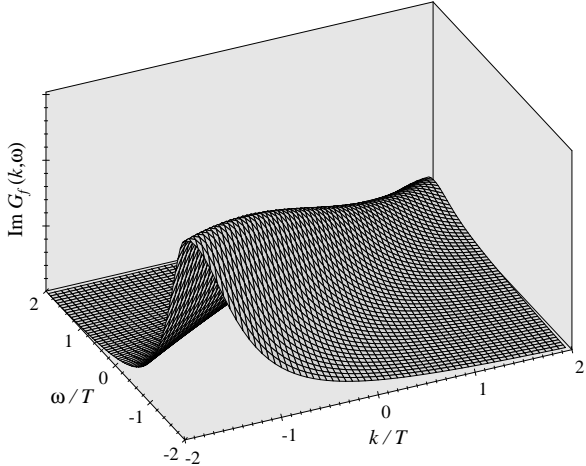}}
\caption{
Low-frequency photoemission intensity near a nodal point,
${\rm Im}\,G_f(k,\omega)\,n_f(\omega)$, given by (\ref{conc1}), (\ref{conc2}).
Here, $\Gamma_f/T=0.58$, which is the result for the $d$-wave to $(s^{*}+id)$-wave
or $(d_{x^2-y^2}+id_{xy})$-wave
transition discussed in Sec.~\ref{sec:dsid},
and $n_f(\omega) = [\exp(\omega/T)+1]^{-1}$ denotes the Fermi function.
}
\label{dampf1}
\end{figure}
For small $\omega,k$, our result for $\widetilde{\Sigma}_f$
in (\ref{dampr}) is
\begin{equation}
\widetilde{\Sigma}_f = i \Gamma_f~~;~~~\mbox{$|\omega|$ and $|k|<T$},
\label{conc2}
\end{equation}
where $\Gamma_f/T$ is a universal number. For the $d$-wave to
$(s^{*}+id)$-wave or $(d_{x^2 - y^2} + i d_{xy})$-wave
transition we estimated $\Gamma_f/T = 0.58$ in
Section~\ref{sec:damp}, while for the onset of a certain type of
${\cal C}$ ordering in a $d$-wave superconductor we obtained
$\Gamma_f /T = 1.35$ in Section~\ref{sec:cd}.
We plot the results (\ref{conc1},\ref{conc2}) in Fig.~\ref{dampf1}.

\begin{figure}
\epsfxsize=3.1in
\centerline{\epsffile{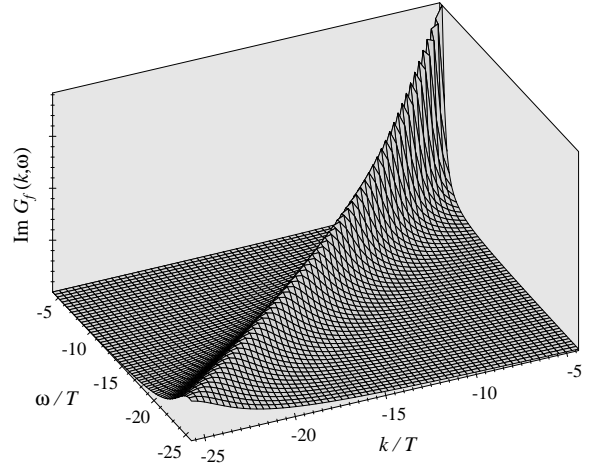}}
\caption{
High-frequency photoemission intensity near a nodal point,
${\rm Im}\,G_f(k,\omega)\,n_f(\omega)$, given by
(\ref{conc3}), (\ref{conc4}), with a small impurity
contribution $U_{\text{imp}}^2=0.1$.
}
\label{dampf2}
\end{figure}

For large $\omega$ or $k$, our result is in (\ref{dsid17}).
For small $\eta_f$, this can be written as
\begin{eqnarray}
&~& \widetilde{\Sigma}_f = \frac{\eta_f}{2} \bigg[ (k-\omega)
\left(\ln \left( \frac{|\omega^2-k^2|}{T^2} \right) - 1 \right)
\nonumber \\ &+& i \pi |\omega-k| \theta(|\omega|-|k|)
\bigg]~;~\mbox{$|\omega|$ or $|k| \gg T$}, \label{conc3}
\end{eqnarray}
where $\theta$ is the unit step function.
Note that the imaginary part of $\widetilde{\Sigma}_f$ vanishes
for $|\omega| \leq |k|$, and, in the present form, this will lead
to an infinite spectral density at the threshold $|\omega| = |k|$.
However, this is repaired by considering corrections to
(\ref{conc3}). Within the scaling limit of the universal theories
being considered here, we evaluate the expression (\ref{d3}) in
Appendix~\ref{app:damp}; at
$T>0$, but with $|\omega|,|k| \gg T$, we obtain in addition to the
leading term in (\ref{conc3}), subleading $T$-dependent
corrections $\mbox{Re} \widetilde{\Sigma}_f^{(1)} \sim T^2/\omega$
and $\mbox{Im} \widetilde{\Sigma}_f^{(1)} \sim T^3/\omega^2$.
These are still very small contributions, and so we can expect
that the system will be exceptionally sensitive to non-universal
corrections to scaling right at the threshold frequency. We
believe that the most important of these will come from elastic
scattering off impurities; for a weak impurity scattering
potential, $U_{\text{imp}}$, we have the additional contribution
\begin{equation}
\Sigma_{\text{imp}} \sim i U_{\text{imp}}^2 |\omega|
\,.
\label{conc4}
\end{equation}
We have added (\ref{conc4}), with a very small prefactor,
to (\ref{conc3}) and plotted the
result in Fig.~\ref{dampf2}; the contribution of (\ref{conc4}) can
be neglected almost everywhere except right near the threshold.

\begin{figure}
\epsfxsize=3.1in
\centerline{\epsffile{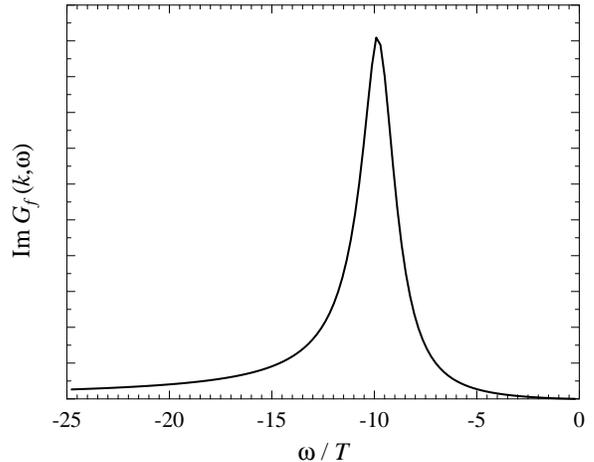}}
\caption{
Sample spectral function for $k/T=10$ -- this is
a cut through the spectrum of Fig.~\protect\ref{dampf2}.
The asymmetry of the lineshape is clearly visible
(and it is not simply due to the Fermi function occurring
as prefactor of the photoemission intensity).
}
\label{dampf3}
\end{figure}

It is interesting to note that our results (\ref{conc2}) and
(\ref{conc3}) bear a superficial similarity to the ``marginal
Fermi liquid'' fitting functions \cite{varma}.
More specifically, ({\em i\/}) in the latter approach,
the prefactor of the power of $T$ in
(\ref{conc1}) is absent; ({\em ii\/}) the small $|\omega|/T$
behavior of the self-energy in (\ref{conc2}) is similar to that of
the marginal Fermi liquid; ({\em iii\/}) for large $|\omega|/T$,
the $k$ dependence in (\ref{conc3}) (as in $\mbox{Im} \widetilde{\Sigma}
\sim |\omega-k| \theta (|\omega|-|k|)$)
is replaced simply by $\mbox{Im} \widetilde{\Sigma} \sim
|\omega|$. A significance consequence of this last difference
is that our large $k$ spectral densities have more asymmetric
lineshapes (even before being multiplied by the Fermi function,
as is necessary for photoemission experiments), than those found
in the marginal Fermi liquid functions;
this is illustrated in Fig.~\ref{dampf3}.
It would be interesting for experiments to test for the $k$
dependence predicted in (\ref{conc3}).

Finally, we reiterate an important feature of our analysis
of quantum criticality: we have only found Lorentz-invariant
fixed points at which all excitations have
equal and isotropic velocities. Section~\ref{sec:anisotropy} showed
that, for the $d$-wave to $(s^{*}+id)$-wave or $(d_{x^2-y^2}+i d_{xy})$-wave
transition,
such a fixed point was at least linearly
stable to perturbations which {\em e.g.} set
$v_F \neq v_{\Delta}$.
However, this does not rule out the possibility that
there may be other non-Lorentz-invariant
fixed points of the renormalization group
equations in which $v_F /v_{\Delta}$ is significantly different
from unity.
Experimentally \cite{mesot}, it is clear that $v_{F} \neq
v_{\Delta}$, but this is still compatible with
a Lorentz-invariant fixed
point: we found in Section~\ref{sec:anisotropy}
that the leading irrelevant operator which breaks Lorentz symmetry
had a scaling dimension of very small absolute value
($\epsilon/21 \approx 0.048$), and so the system can reside in a
transient region with $v_F \neq v_{\Delta}$ over a very wide
temperature range.
This issue will be addressed further in future work.

~\\
\emph{Notes added:} ({\em i\/}) Recent THz conductivity measurements
on ${\rm Bi}_2 {\rm Sr}_2 {\rm Ca Cu}_2 {\rm O}_{8+\delta}$ by
Corson {\em et al.} \cite{corson2} have obtained a quasiparticle
relaxation rate linearly proportional to $T$, at temperatures well
below $T_c$. Combined with the photoemission experiments
\cite{valla}, these results provide strong support for quantum
critical damping of the nodal quasiparticles in the $d$-wave
superconductor, as is expected near a class A quantum critical
point between two superconducting states. Corson {\em et al.}
observe the quantum critical damping above 20K, which suggests
that the energy per coherence volume of the second superconducting
state (say, the $(d_{x^2-y^2} + i d_{xy})$
superconductor) is higher than that of the $d$-wave superconductor
by less than 20 K.
\newline
({\em ii\/}) A recent work\cite{newton} has given a unified discussion of the
quantum phase transitions considered here, and those involving
${\cal M}$ symmetry breaking considered in Ref.~\onlinecite{qimp}.


\acknowledgements

We thank L.~Balents, A.~Castro Neto,
D.~Khomskii, J.~Kirtley, S.~Kivelson,
J.~B.~Marston, C.~Nayak, C.~M.~Varma and especially P.~Johnson,
M.~Norman, M.~Randeria, and J.~Zaanen
for useful discussions. This research was supported by US NSF
Grant No DMR 96--23181 and by the DFG (VO 794/1-1).


\appendix
\section{Perturbation theory for fermion damping}
\label{app:damp}
This appendix will compute the $T>0$ scaling function for the fermion
spectral weight in (\ref{i1}) for the case of a quantum critical
point between a $d$-wave and a $(s^{*}+id)$-wave or $(d_{x^2-y^2}+i d_{xy})$-wave
superconductor.
We will use a simple renormalized perturbative expansion at the
fixed point found in Section~\ref{sec:dsid}. We will show that
such a procedure leads to spurious singularities in the low
frequency regime $\hbar \omega < k_B T$. These singularities were
cured by the self-consistent analysis described in
Section~\ref{sec:damp}.

 From the expression (\ref{d3}), we obtain to leading order in $\lambda^2$
and $\epsilon$
\begin{eqnarray}
&& \left( \frac{\mu}{T} \right)^{-\eta_f} G_f^{-1} (k, \omega_n) =
\nonumber \\
&&~
(-i \omega_n + k_x \tau^z + k_y \tau^x )\left( 1 -
\frac{\lambda^{\ast 2}}{2} \left(\frac{1}{\epsilon} + \ln(\mu/T)
\right)\right) \nonumber \\
&&~+ \frac{\lambda^{\ast 2}}{S_{d+1}} \int \!\! \frac{d^d p}{(2
\pi)^d} T \sum_{\epsilon_n} \frac{- i \epsilon_n + p_x \tau^z +
p_y \tau^x}{(p^2 + \epsilon_n^2)((k-p)^2 + (\omega_n -
\epsilon_n)^2)}.
\nonumber
\end{eqnarray}
Evaluating the frequency summation, and performing the momentum
integration over the terms not involving any thermal Bose or Fermi
factors, we obtain
\begin{eqnarray}
&& \left( \frac{\mu}{T} \right)^{-\eta_f} G_f^{-1} (k, \omega_n) =
-\widetilde{\Sigma}_f^{(1)} (k, \omega_n) +
\nonumber \\
&&~
(-i \omega_n + k_x \tau^z + k_y \tau^x )\left( 1 -
\frac{\eta_f}{2} \left(\ln\left( \frac{\omega_n^2 + k^2}{e T^2}
\right)
\right)\right) \nonumber \\
&&~
\end{eqnarray}
where we have used (\ref{dsid14}), and $\widetilde{\Sigma}_f^{(1)}$
is a thermal contribution which vanishes as $T \rightarrow 0$.
To leading order in $\epsilon$, the expression for $\widetilde{\Sigma}_f^{(1)}$
can be evaluated in $d=3$, and we obtain for $k$ along the $x$
direction (recall that below (\ref{dsid1}) we rotated the axes by
45 degress from the axes of the square lattice):
\begin{eqnarray}
&& \widetilde{\Sigma}_f^{(1)} (k, \omega_n) = -
16 \pi^2 \eta_f \int \frac{d^3 p}{(2 \pi)^3}
\frac{1}{2 p |{\bf p} - {\bf k}|} \nonumber \\
&&~ \times \left(
\frac{[n(|{\bf p} - {\bf k}|)+f(p)][(p_x \tau^z (p-|{\bf p} - {\bf k}|)
- i \omega_n p]}{(p - |{\bf p} - {\bf k}|)^2 + \omega_n^2} \right.\nonumber\\
&& \!\!\!
\left. + \frac{[n(|{\bf p} - {\bf k}|)-f(p)][p_x \tau^z (p+|{\bf p} - {\bf k}|)
- i \omega_n p]}{(p + |{\bf p} - {\bf k}|)^2 + \omega_n^2} \right)
\label{app2}
\end{eqnarray}
where $p = |{\bf p}|$, $k = |{\bf k}|$, $n(k)$ is the Bose function,
and $f(k)$ is the Fermi function.
The expression in
(\ref{app2}) is reliable for $|\omega|, k \gg T$: evaluating the integrals
in this regime we obtain the estimates
quoted below (\ref{conc3}). On the contrary, for $|\omega|, k \ll
T$, the above expressions are pathological; we obtain {\em e.g.}
$\mbox{Im} \widetilde{\Sigma}_f^{(1)} (0, \omega) \sim T^2
\delta(\omega)$. This should be contrasted with the smooth
behavior as a function of $\omega$ assumed in (\ref{dampr}). The
latter is the correct result on physical grounds \cite{SY}, and estimation
of the damping constants requires a self-consistent approach like
that followed in Section~\ref{sec:damp}.



\begin{thebibliography}{}

\bibitem{zaanen2} J.~Zaanen, Physica C {\bf 317}, 217 (1999).

\bibitem{bondsite} J.~Tworzydlo, O.~Y.~Osman, C.~N.~A.~van Duin, and J.~Zaanen,
\prb {\bf 59}, 115 (1999).

\bibitem{pnas}
V.~J.~Emery, S.~A.~Kivelson, and J.~M.~Tranquada,
Proc. Natl. Acad. Sci. USA {\bf 96},
8814 (1999).

\bibitem{sciencereview} See the review S. Sachdev, Science {\bf
288}, 475 (2000).

\bibitem{sr} S.~Sachdev and N.~Read, Int. J. Mod. Phys. B
{\bf 5}, 219 (1991).

\bibitem{stripeprl} M.~Vojta and S.~Sachdev, \prl {\bf 83}, 3916 (1999).

\bibitem{rs0} N.~Read and S.~Sachdev, Phys. Rev. Lett. {\bf 62},
1694 (1989); Phys. Rev. B {\bf 42}, 4568 (1990).

\bibitem{gsh} M.~P.~Gelfand, R.~R.~P.~Singh, and D.~A.~Huse, \prb
{\bf 40}, 10801 (1989).

\bibitem{rs1} N.~Read and S.~Sachdev, Phys. Rev. Lett. {\bf 66},
1773 (1991).

\bibitem{kotov} V.~N.~Kotov, J.~Oitmaa, O.~P.~Sushkov, and Zheng
Weihong, \prb {\bf 60}, 14613 (1999);
R.~R.~P.~Singh,~Zheng Weihong,~C.~J.~Hamer, and J.~Oitmaa, \prb {\bf 60}, 7278 (1999);
V.~N.~Kotov and O.~P. Sushkov, \prb {\bf 61}, 11820 (2000).

\bibitem{sushkov} O.~P.~Sushkov,
cond-mat/9907400,
cond-mat/0002421.

\bibitem{fn1} We define the operator which creates a particle on a
bond by $a^{\dagger}_{\alpha} = (c_{1\alpha}^{\dagger} + c_{2
\alpha}^{\dagger})/\sqrt{2}$, where $1,2$ are sites on the ends of
the bond; then the bond charge density is $a^{\dagger} a$.
Clearly, there can be no bond-charge density modulation in spin models
in which
the electron occupation number is constrained to be exactly unity
per site. However, it appears once the connection to the
operators of the parent Hubbard-like model is accounted for.

\bibitem{book} S.~Sachdev, {\em
{Quantum Phase Transitions}}, Cambridge University Press, Cambridge (1999).

\bibitem{affmar} I.~Affleck and J.~B.~Marston, \prb {\bf 37}, 3774
(1988).

\bibitem{schulzoaf} H.~Schulz, Phys. Rev. B {\bf 39}, 2940 (1989).

\bibitem{biq} J.~B.~Marston and I.~Affleck, \prb {\bf 39}, 11538 (1989).

\bibitem{wang} Z.~Wang, G.~Kotliar, X.-F.~Wang, Phys. Rev. B
{\bf 42}, 8690 (1990).

\bibitem{nerses} A.~Nersesyan, Phys. Lett. A {\bf 153}, 49 (1991);
A.~Nersesyan, G.~Japaridze, and I.~Kimeridze, J.~Phys.: Condens.
Matter {\bf 3}, 3353 (1991).

\bibitem{kfe} S.~A.~Kivelson, E.~Fradkin, and V.~J.~Emery, Nature
{\bf 393}, 550 (1998). See also
V.~J.~Emery, S.~A.~Kivelson, and O.~Zachar, Phys.
Rev. B {\bf 56}, 6120 (1997) and Ref~\protect\onlinecite{ekl}.

\bibitem{zaanen} J.~Zaanen and O.~Gunnarsson, Phys. Rev. B {\bf 40},
7391 (1989).

\bibitem{schulz} H.~Schulz, J. de Physique {\bf 50}, 2833 (1989).

\bibitem{machida} K.~Machida, Physica {\bf 158C}, 192 (1989);
M.~Kato, K.~Machida, H.~Nakanishi, M.~Fujita, J. Phys. Soc. Jpn
{\bf 59}, 1047 (1990).

\bibitem{shraiman} B.~I.~Shraiman and E.~D.~Siggia, Phys. Rev. B
{\bf 42}, 2485 (1990); S.~Sachdev, Phys. Rev. B {\bf 49}, 6770
(1994).

\bibitem{zachar} O.~Zachar, S.~A.~Kivelson, and V.~J.~Emery,
\prb {\bf 57}, 1422 (1998).

\bibitem{collinear} J.~M.~Tranquada, J.~D.~Axe, N.~Ichikawa, Y.~Nakamura,
S.~Uchida, and B.~Nachumi, Phys. Rev. B {\bf 54}, 7489 (1996).

\bibitem{shirane} H.~Kimura, K.~Hirota, H.~Matsushita, K.~Yamada, Y.~Endoh,
S.-H.~Lee, C.~F.~Majkrzak, R.~Erwin, G.~Shirane, M.~Greven, Y.~S.~Lee, M.~A.~Kastner,
and R.~J.~Birgeneau, \prb {\bf 59}, 6517 (1999).

\bibitem{younglee} Y.~S.~Lee, R.~J.~Birgeneau, M.~A.~Kastner, Y.~Endoh,
S.~Wakimoto, K.~Yamada, R.~W.~Erwin, S.-H.~Lee, and
G.~Shirane, \prb {\bf 60}, 3643 (1999).

\bibitem{castro} A.~H.~Castro Neto and D.~Hone, \prl {\bf 76},
2165 (1996).

\bibitem{qimp} M.~Vojta, C.~Buragohain, and S.~Sachdev, Phys. Rev.
B {\bf 61}, 15152 (2000).

\bibitem{imai} A.~W.~Hunt, P.~M.~Singer,
K.~R.~Thurber, and T.~Imai, Phys. Rev. Lett. {\bf 82},
4300 (1999);  T.~Imai, C.~P.~Slichter, K~Yoshimura, and
K.~Kosuge, {\em ibid} {\bf 70}, 1002 (1993);
T.~Imai, C.~P.~Slichter, K.~Yoshimura,
M.~Katoh, and K.~Kosuge, \prl {\bf 71}, 1254 (1993);
S.~Fujiyama, M.~Takigawa, Y.~Ueda, T.~Suzuki, N.~Yamada, \prb {\bf 60}, 9801 (1999).

\bibitem{morr} D.~K.~Morr, J.~Schmalian, and D.~Pines,
cond-mat/0002164.

\bibitem{gabi} G.~Kotliar, Phys. Rev. B {\bf 37}, 3664 (1988).

\bibitem{did1} D.~S.~Rokhsar, Phys. Rev. Lett {\bf 70}, 493 (1993).

\bibitem{did2}
R.~B.~Laughlin, Physica {\bf 243C}, 280 (1994);
R.~B.~Laughlin, Phys. Rev. Lett. {\bf 80}, 5188 (1998);
A.~V.~Balatsky, \prb {\bf 61}, 6940 (2000);
T.~Maitra,
cond-mat/0002114.

\bibitem{senthil} T.~Senthil,
J.~B.~Marston, and M.~P.~A.~Fisher, Phys. Rev. B {\bf 60}, 4245
(1999).

\bibitem{ws} S.~R.~White and D.~J.~Scalapino, Phys. Rev. Lett.
{\bf 80}, 1272 (1998); {\bf 81}, 3227 (1998);
\prb {\bf 60}, R753 (1999).                     

\bibitem{honer} C.~Honerkamp, M.~Salmhofer, N.~Furukawa, and T.~M.~Rice,
cond-mat/9912358.

\bibitem{leder} U.~Ledermann, K.~Le Hur, and T.~M.~Rice,
cond-mat/0002445.

\bibitem{stojk} B.~P.~Stojkovi\'{c}, Z.~G.~Yu, A.~R.~Bishop,
A.~H.~Castro Neto, and N.~Gr{\o}nbech-Jensen, \prl {\bf 82}, 4679 (1999);
cond-mat/9911380.

\bibitem{valla} T.~Valla, A.~V.~Fedorov, P.~D.~Johnson,
B.~O.~Wells, S.~L.~Hulbert, Q.~Li, G.~D.~Gu, and N.~Koshizuka, Science {\bf
285}, 2110 (1999).

\bibitem{clap} A.~V.~Balatsky, P.~Kumar, and J.~R.~Schrieffer,
Phys. Rev. Lett. {\bf 84}, 4445 (2000).
Our order parameter, $\phi$, measures the amplitude
oscillations of the $d_{xy}$ component about zero amplitude
at a fixed phase ($\pm \pi/2$)
relative to the stable $d_{x^2-y^2}$ state, whereas Balatsky {\em et
al.} work in the state with global $(d_{x^2-y^2} + i d_{xy})$
order and consider oscillations in the relative phase between the
two components, both of which have a non-zero amplitude. The
frequency of the amplitude oscillations we consider
vanishes at the critical point between the two superconductors.
Also, the susceptibility corresponding to this amplitude mode
will diverge at this critical point.

\bibitem{didexp} K.~Krishana, N.~P.~Ong, Q.~Li, G.~D.~Gu, and N.~Koshizuka,
Science {\bf 277}, 83 (1997);
H.~Aubin, K.~Behnia, S.~Ooi, and T.~Tamegai, \prl {\bf 82}, 624 (1999).

\bibitem{balents} L.~Balents, M.~P.~A.~Fisher, and C.~Nayak, Int. J. Mod.
Phys. B {\bf 12}, 1033 (1998).

\bibitem{castellani} C.~Castellani, C.~di Castro, and M.~Grilli,
\prl {\bf 75}, 4650 (1995), Z. Phys. B {\bf 103}, 137 (1997),
J. Phys. Chem. Solids {\bf 59}, 1694 (1998).

\bibitem{shen} Z.-X.~Shen and D.~S.~Dessau, Phys. Rep. {\bf 253}, 1 (1995).

\bibitem{campu} J.~C.~Campuzano {\em et al.}, in
{\em The Gap Symmetry and Fluctuations in High-$T_{c}$ Superconductors},
eds. J. Bok {\em et al.} (Plenum, New York, 1998), p. 229.

\bibitem{fed} A.~V.~Fedorov, T.~Valla, P.~D.~Johnson,
Q.~Li, G.~D.~Gu, and N.~Koshizuka, Phys. Rev. Lett. {\bf 82}, 2179 (1999).

\bibitem{mesot} J.~Mesot, M.~R.~Norman, H.~Ding, M.~Randeria, J.~C.~Campuzano,
A.~Paramekanti, H.~M.~Fretwell, A.~Kaminski, T.~Takeuchi, T.~Yokoya,
T.~Sato, T.~Takahashi, T.~Mochiku, and K.~Kadowaki, \prl {\bf 83}, 840
(1999).

\bibitem{mesot2} A.~Kaminski, J.~Mesot, H.~Fretwell, J.~C.~Campuzano,
M.~R.~Norman, M.~Randeria, H.~Ding, T.~Sato, T.~Takahashi, T.~Mochiku,
K.~Kadowaki, and H.~Hoechst, \prl {\bf 84}, 1788
(2000).

\bibitem{kirtley} F.~Tafuri and J.~R.~Kirtley,
cond-mat/0003106.

\bibitem{sigrist} D.~B.~Bailey, M.~Sigrist, and R.~B.~Laughlin,
\prb {\bf 55}, 15239 (1997); M.~Sigrist, Prog. Theor. Phys.
{\bf 99}, 899 (1998).

\bibitem{ekl} V.~J.~Emery, S.~A.~Kivelson, and H.~Q.~Lin, Phys.
Rev. Lett. {\bf 64}, 475 (1990); Phys. Rev. B {\bf 42}, 6523 (1990).

\bibitem{sk1} S.~A.~Kivelson and V.~J.~Emery in {\it Strongly
Correlated
Electronic Materials:  The Los Alamos
Symposium 1993}, edited by K.S. Bedell,
Z. Wang, D.E. Meltzer, A.V. Balatsky, and E. Abrahams,
(Addison-Wesley, Reading, Massachusetts, 1994) p. 619.

\bibitem{sk2} U.~L\"ow, V.~J.~Emery, K.~Fabricius, and
S.~A.~Kivelson,
Phys. Rev. Lett. {\bf 72}, 1918  (1994).

\bibitem{hm} C.~Hellberg and E.~Manousakis, Phys. Rev. Lett. {\bf
78}, 4069 (1997).

\bibitem{sstri} S.~Sachdev, \prb {\bf 45}, 12377 (1992).

\bibitem{ziqiang} S.~Sachdev and Z.~Wang, \prb {\bf 43}, 10229
(1991).

\bibitem{foerster} D.~Foerster,
cond-mat/0001385.

\bibitem{rsnp} N.~Read and S.~Sachdev, Nucl. Phys. B {\bf 316},
609 (1989); D.~Rokhsar, Phys. Rev. B {\bf 42}, 2526 (1990).

\bibitem{grilli} M.~Grilli, C.~Castellani, and G.~Kotliar, Phys.
Rev. B {\bf 45}, 10805 (1992).

\bibitem{coulnote}
In our earlier work \cite{stripeprl} we had implicitly assumed
a cut-off or exponential decay of $V_{ij}$ for large distances.
Doing so, the ``striped'' states survive down to $\delta\to 0$,
and the relation $p \sim 1/\delta$ is exact in this limit.

\bibitem{jtran}
J.~M.~Tranquada, J. Phys. Chem. Solids {\bf 59}, 2150 (1998).

\bibitem{birg} S.~Wakimoto, R.~J.~Birgeneau, Y.~Endoh, P.~M.~Gehring,
K.~Hirota, M.~A.~Kastner, S.~H.~Lee, Y.~S.~Lee, G.~Shirane, S.~Ueki,
and K.~Yamada, \prb {\bf 60}, R769 (1999);
S.~Wakimoto, K.~Yamada, S.~Ueki, G.~Shirane, Y.~S.~Lee,
S.~H.~Lee, M.~A.~Kastner, K.~Hirota, P.~M.~Gehring, Y.~Endoh, R.~J.~Birgeneau,
J. Phys. Chem. Solids {\bf 60}, 1079 (1999).

\bibitem{mpaf} H.-H.~Lin, L.~Balents, and M.~P.~A.~Fisher, Phys.
Rev. B {\bf 56}, 6569 (1997).

\bibitem{uchida} N.~Ichikawa, S.~Uchida, J.~M.~Tranquada,
T.~Niemoeller, P.~M.~Gehring, S.-H.~Lee, and J.~R.~Schneider,
cond-mat/9910037.

\bibitem{julich} G~B.~Teitel'baum, B.~B\"uchner, and
H.~de~Gronckel, \prl {\bf 84}, 2949 (2000).

\bibitem{bradlec} J.~B.~Marston,
cond-mat/9904437,
proceedings of {\it The International Workshop on Superconductivity,
Magneto-Resistive Materials, and Strongly Correlated Systems},
Hanoi, Vietnam, January 1999 (to appear).

\bibitem{ivanov} D.~A.~Ivanov, P.~A.~Lee, X.-G.~Wen,
Phys. Rev. Lett. {\bf 84}, 3958 (2000);
X.~G.~Wen and P.~A.~Lee, {\it Phys. Rev. Lett.} {\bf 76},
503 (1996); P.~A.~Lee, N.~Nagaosa, T.~K.~Ng, X.-G.~Wen, {\it Phys.
Rev. B} {\bf 57}, 6003 (1998).

\bibitem{nayak} C.~Nayak,
cond-mat/0001303;
cond-mat/0001428.

\bibitem{KotLiu} B.~G.~Kotliar and
J.~Liu, Phys. Rev. B {\bf 38}, 5142 (1988).

\bibitem{zeyer}  E.~Cappelluti and R.~Zeyer, \prb {\bf 59}, 6475 (1999).

\bibitem{sunstripes} J.~B.~Marston and C.-H.~Chung, to be published.

\bibitem{dunghai} D.-H.~Lee, \prb {\bf 60}, 12429 (1999).

\bibitem{zj} J.~Zinn-Justin, {\em Quantum Field Theory and
Critical Phenomena}, Clarendon Press, Oxford (1993), Section 10.8.

\bibitem{ssepsilon} S.~Sachdev, Phys. Rev. B {\bf 55}, 142 (1997);
Phys. Rev. B {\bf 59}, 14054 (1999).

\bibitem{note}
In Ref~\protect\onlinecite{book} a more general fitting
function was used for $G_b$, equivalent to the form
$G_b^{-1} \propto ( \alpha^2 k^2 - \omega^2 - i \omega \omega_1 +
\omega_2^2)$; a very good best-fit to the exact result was found
for $\alpha = 1.02$ and $\omega_2 = 2.01 \omega_1$
(see Eqn (4.116) in Ref~\protect\onlinecite{book}). The form in
(\protect\ref{dampr}) is equivalent to assuming $\alpha=1$ and
$\omega_2 = 2 \omega_1$, which are clearly excellent
approximations.

\bibitem{bgz} E.~Br\'{e}zin, J.~C.~Le Guillou, J.~Zinn-Justin in
{\em Phase Transitions and Critical Phenomena}, {\bf 6}, C.~Domb
and M.~S.~Green eds, Academic Press, London (1976).

\bibitem{bza} G.~Baskaran, Z.~Zou, and P.~W.~Anderson,
Solid State Commun. {\bf 63}, 973 (1987);
A.~E.~Ruckenstein, P.~J.~Hirschfeld, and J.~Appel,
Phys. Rev. B {\bf 36}, 857 (1987).

\bibitem{cmv} C.~M.~Varma, Phys. Rev. B {\bf 55}, 14554 (1997);
Phys. Rev. Lett. {\bf 83}, 3538 (1999);
Phys. Rev. B {\bf 61}, R3804 (2000).

\bibitem{phase} V.~P.~Gusynin, V.~M.~Loktev, and S.~G.~Sharapov,
Sov. Phys. JETP {\bf 90}, 993 (2000);
M.~Franz and A.~J.~Millis, Phys. Rev. B {\bf 58}, 14572 (1998);
H.-J.~Kown and A.~T.~Dorsey, Phys. Rev. B {\bf 59}, 6438 (1999).

\bibitem{varma} C.~M.~Varma, P.~B.~Littlewood, S.~Schmitt-Rink
and A.~E.~Ruckenstein, Phys. Rev. Lett. {\bf 63}, 1996 (1989);
E.~Abrahams and C.~M.~Varma,
Proc. Natl. Acad. Sci. USA {\bf 97}, 5714 (2000).

\bibitem{corson2} J.~Corson, J.~Orenstein, and J.~N.~Eckstein,
cond-mat/0003243.

\bibitem{newton} S.~Sachdev and M.~Vojta, cond-mat/0005250.

\bibitem{SY} S.~Sachdev and J.~Ye, Phys. Rev. Lett. {\bf 69}, 2411
(1992).


\end{thebibliography}
\end{document}